\documentclass[10pt, english]{article}
\usepackage{sectsty}
\sectionfont{\fontsize{12}{10}\selectfont}
\usepackage{geometry}
\usepackage{setspace}
\usepackage{babel}
\usepackage{xcolor}
\usepackage{graphicx}
\usepackage{amsmath,amsfonts}
\usepackage{xcolor}
\usepackage{graphicx,indentfirst,subfigure,epsfig}
 \usepackage{varioref}
 \usepackage{wrapfig}
 \usepackage{subfigure}
 \usepackage{subfigmat}
 \usepackage{amsthm}
\usepackage{hyperref}
\hypersetup{colorlinks, citecolor=red, linkcolor=black, urlcolor=red, breaklinks=true}
\usepackage{scrextend}
\usepackage{nomencl}
\makenomenclature
\usepackage{booktabs}
\usepackage{longtable}
\usepackage{soul}

\usepackage[italic]{mathastext}
\usepackage{bm}
  \def\clap#1{\hbox to 0pt{\hss#1\hss}}

\providecommand{\mat}[1]{\bm{#1}}%
\renewcommand{\vec}[1]{\mathbf{#1}}

\providecommand{\mH}{\ensuremath{\mat{H}}}
\providecommand{\mI}{\ensuremath{\mat{I}}}

\providecommand{\mK}{\mathbf{K}}

\providecommand{\mR}{\ensuremath{\mat{R}}}

\providecommand{\mX}{\ensuremath{\mathbf{X}}}

\providecommand{\vd}{\ensuremath{\vec{d}}}

\providecommand{\vf}{\ensuremath{\vec{f}}}

\providecommand{\vm}{\ensuremath{\vec{m}}}

\providecommand{\vr}{\ensuremath{\vec{r}}}

\providecommand{\vt}{\ensuremath{\vec{t}}}

\providecommand{\vv}{\ensuremath{\vec{v}}}
\providecommand{\vw}{\ensuremath{\vec{w}}}
\providecommand{\vx}{\ensuremath{\vec{x}}}

\providecommand{\vz}{\ensuremath{\vec{z}}}

\usepackage{fancyhdr}
\begin{document}

\title{\small{\textbf{SPATIAL ANOMALY DETECTION WITH OPTIMAL TRANSPORT}}}
\author{\small{Pranay Seshadri$^{\dagger}$\thanks{Address all correspondence to p.seshadri@imperial.ac.uk.}, Andrew B. Duncan$^\dagger$, George Thorne$^{\ddagger}$}, and  Ra\'{u}l V\'{a}zquez D\'{i}az$^{\ddagger}$}
\date{\small{$^{\dagger}$Imperial College London, London, United Kingdom.}\\\small{$^{\ddagger}$Rolls-Royce plc, Derby, United Kingdom.}}

\maketitle 
\pagestyle{fancy}
\begin{abstract}
\small{This manuscript outlines an automated anomaly detection framework for jet engines. It is tailored for identifying spatial anomalies in steady-state temperature measurements at various axial stations in an engine. The framework rests upon ideas from optimal transport theory for Gaussian measures which yields analytical solutions for both Wasserstein distances and barycenters. The anomaly detection framework proposed builds upon our prior efforts that view the spatial distribution of temperature as a Gaussian random field. We demonstrate the utility of our approach by training on a dataset from one engine family, and applying them across a fleet of engines---successfully detecting anomalies while avoiding both false positives and false negatives. Although the primary application considered in this paper are the temperature measurements in engines, applications to other internal flows and related thermodynamic quantities are made lucid.}
\end{abstract}

\maketitle 

\tableofcontents

\section{Introduction}
\label{sec:intro}
To discern if an observed measurement is anomalous, one must have some baseline measurement to compare it against. This introduces two requirements. First, a metric that measures the difference---or more generally the \emph{distance}---between the observed and baseline. Second, a \emph{threshold} that delineates whether the distance is large enough to be classified as anomalous or not. For instance, if the \emph{absolute value of the difference} between the observed measurement and the baseline is \emph{greater than 2}, then it is anomalous. 

This delineation between an \emph{acceptable} and anomalous observation is tedious to quantify when there are multiple \emph{related} observations and consequently multiple related baseline measurements. Such is the case that we consider in this paper. More specifically, we wish to identify spatial anomalies in stagnation temperature sensors in an engine; applications to other thermodynamic quantities, and indeed other internal flow applications are extensions of the present work. The sensors considered in this work are positioned on both rakes and vanes, and thus at a given axial station are functions of radial and circumferential locations $\left(r, \theta\right)$. Rake placements may differ across engines, as may absolute values of their thermodynamic quantities. Thus, one cannot simply compute the distance between two sets of measurements. Moreover, in an operational engine environment, practitioners want to know if sensors are reporting anomalous values, and if so, which ones. Thus, a scalar distance between an observed and baseline set of measurements, in isolation, will fail to offer necessary information on the precise location of the anomaly and potential cause thereof. Finally, to arrive at such a delineation, data-driven anomaly detection methods \cite{hipple2020using, sepe2021physics, xu2016bayesian, yan2019accurate, zhao2016review} are seen as the way forward, with the caveat that they require large training repositories. This may be infeasible for certain applications, such as ours, where anomaly detection over a higher granularity of measurements is sought for which training data is limited by virtue of the costs of well-instrumented engine tests.

To address these issues, in this paper, we consider the following ideas.
\begin{enumerate}
    \item It may be beneficial to construct a probabilistic spatial model using each set of measurements independently. In doing so, we have at hand an annular model for observed measurements, and another annular model for the baseline measurements. At the sensor locations, we expect this probabilistic model to have a very small uncertainty---dictated by the measurement apparatus. As we move away from the sensor locations, the model uncertainty will increase based on model assumptions and data availability.
    \item Prior to computing any distances, it will be important to normalise the data as different engine tests may have relatively higher or lower values and we would not want the distance to be dominated by the apparent difference in the mean. One way to do this would be to normalise by the area average. That said, this area average should not be based on the sensor positions---which would invariably introduce a bias---but rather based on spatially integrating the aforementioned probabilistic model.
    \item It may be prudent to have sufficient granularity for identifying which part of the space the anomaly originates from. Therefore it makes sense to consider multiple anomaly detection tasks. To do this, a vector of distances based on the location of the sensors may be appropriate.  This will yield the location of the spatial anomaly, and offer relative comparisons between neighboring measurements. 
    \item Finally, rather than rely on the availability of a large training repository, it may be sagacious to combine available data with synthetically generated data to boost the overall training repository size. However, the precise manner this synthetic data is to be generated must be carefully considered.
\end{enumerate}

\begin{figure}[h]
    \centering
    \includegraphics[scale=0.3]{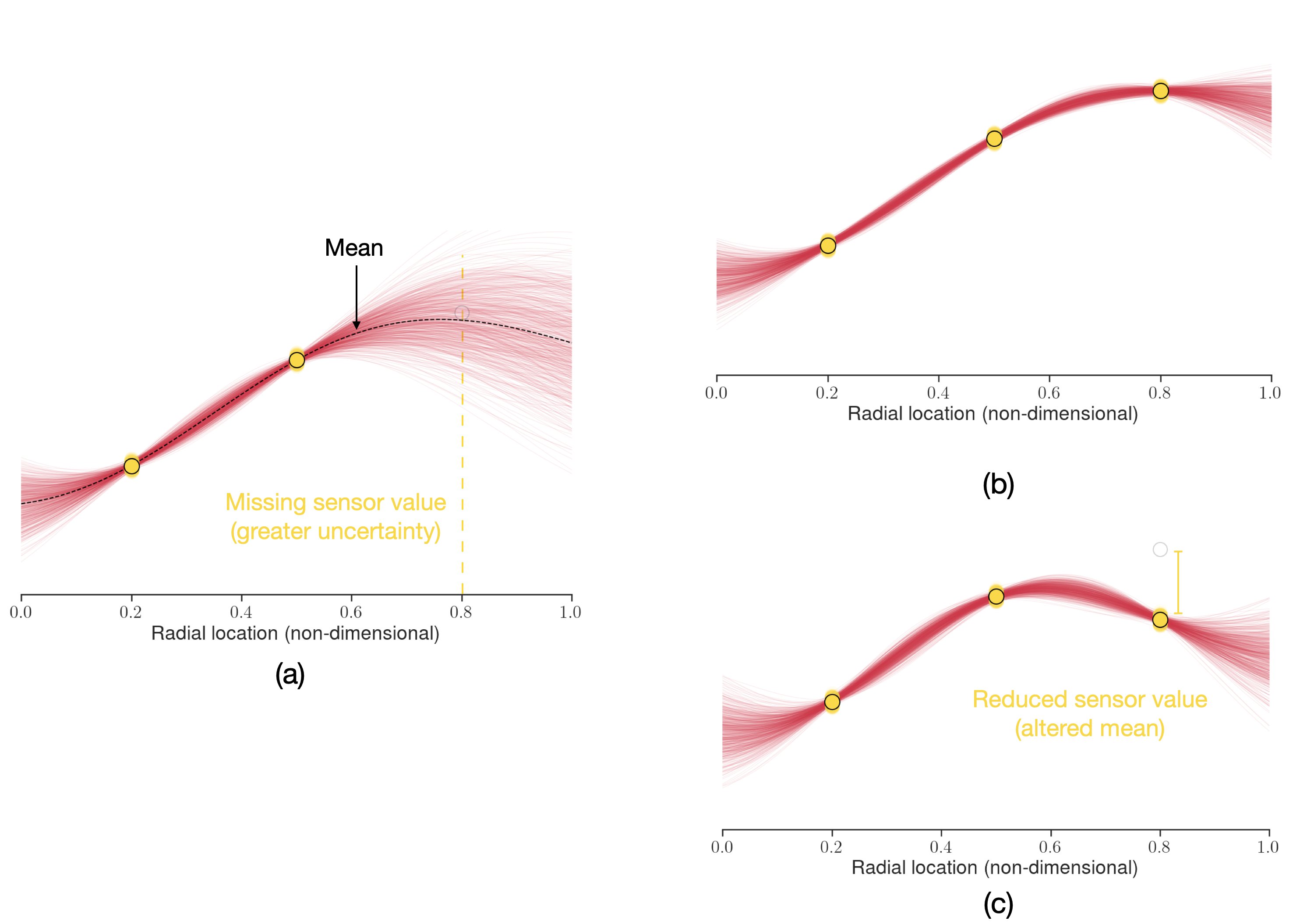}
    \caption{Motivating a probabilistic interpretation of anomaly detection: (a) two sensor measurements; (b) inclusion of an extra sensor; (c) inclusion of an extra sensor with a reduced sensor value.}
    \label{fig:motivation}
\end{figure}

We are still left with the matter of selecting a suitable distance and threshold. In terms of the distance, one point to recognise is that we are no longer comparing scalars or vectors: we are contrasting probability distributions. To motivate this departure from existing anomaly detection literature \cite{chandola2009anomaly}, consider the data shown in Figure~\ref{fig:motivation}. Subfigure (a) shows two sample measurements taken at distinct non-dimensional radial locations. Each measurement is denoted by a circular maker and the numerous circular overlays reflect the uncertainty in a given measurement. The interpolating curves offer plausible explanations of the data, based on any prior information. In a Bayesian context these curves are referred to as the predictive posterior distribution. In (b) we consider the inclusion of an extra sensor that is not observed in (a). It is clear that some of the interpolating curves in (a) are inconsistent with the predictive posterior distribution in (b), in the sense that they lie in the tails of that distribution. From an anomaly detection perspective, we are interested in the following: if the two sensors in (a) represent a baseline (or gold standard) in measurements, then is the measurement at a non-dimensional radial location of 0.8 in (b) anomalous? If our answer only utilised the mean curve in (a), then we would be inclined to say ``no". 

However, we see that curves which would be plausible under the posterior in (a) would become implausible (i.e., lie in the tails of the predictive posterior distribution) for (b). Taking into account this uncertainty we would therefore consider (a) and (b) to be far away from each other. In Figure~\ref{fig:motivation}(c) we illustrate another possible outcome from the sensor's value at 0.8. Now if we assume (b) is the baseline, then we want the difference in the mean to be adequately accounted for. To summarise: in this paper we utilise a probabilistic paradigm for anomaly detection to appropriately account for plausible explanations of the data, which would be consistent with the baseline model and its composite measurements.

Even at the extreme when comparing an observed and baseline, each endowed with only one sensor at the same location $\left(r, \theta\right)$, we are still comparing two probability distributions. This is because the sensor measurement will likely have uncertainty arising from temporal averaging, signal-to-noise filtering, and a variety of thermodynamic calibrations---for converting from volts to Kelvin (or Pascals)---yielding a probability distribution. Thus we restrict our search to distances that can be used to compare probability distributions.

Classical statistical metrics for comparing distributions that may be utilised include the total variation distance, the Hellinger norm, and the $L_2$ (Euclidean) norm. However, these metrics do not consider the underlying space of the distributions. For instance, the distances between two uniform distributions that have similar means is identical to a uniform distribution with a very different one if they all have the same variance. The Kullback-Leibler divergence \cite{kullback1951information}, is possibly a candidate, however it is not a metric in the sense that the distance between a baseline and observed is not equivalent to the distance between the observed and baseline; in other words it is asymmetric\footnote{Note that the Kullback-Leibler divergence can be made symmetric via the Jensen Shannon distance.}. This may introduce additional issues when trying to set a threshold.

What we do therefore is to exploit ideas from the field of optimal transport and Bayesian inference to address the the points raised above (point-by-point respectively). We formalise the ideas discussed above as follows.
\begin{enumerate}
    \item To build a probabilistic spatial model, we use recent ideas in \cite{seshadri2021bayesian, seshadri2021bayesianmass} that view the thermodynamic quantities at an engine axial plane as a Gaussian random field. In scope, this builds upon prior least-squares based methods \cite{seshadri2020spatial, seshadri2020spatialb, lou2021reconstructing}.
    \item To negotiate issues pertaining to normalisation, we compute the Bayesian area average as derived in \cite{seshadri2021bayesian}. This is an analytical calculation as it is a linear operator acting over the Gaussian random field.
    \item We introduce a vector of one-dimensional Wasserstein distances, where each component of the vector is the distance for a particular sensor location's $\left(r, \theta\right)$ coordinate. Our choice in adopting the Wasserstein metric is based on its symmetry, its closed-form expressions for Gaussian distributions, and its ability to factor the underlying space of the distributions.
    \item For generating synthetic data, we introduce a barycentric interpolation methodology that extends standard Wasserstein geodesics---amendable with only two distributions---to a higher dimensional manifold. This permits us to generate synethetic samples that combine multiple baseline datasets. 
\end{enumerate}

Following these introductory comments, the structure of this paper is set down. Section~\ref{sec:gps} offers a cursory overview to Gaussian processes, followed by an overview of the chosen kernel functions and the method for inference. What follows is a condensed form of the fundamentals of optimal transport in section~\ref{sec:opt_transport}, and the spatial anomaly detection framework is detailed in \ref{sec:anomaly} with a specific focus on the distance metric, the threshold selection strategy, and synthetic data generation. Finally, numerical examples of the propose framework at work are given in  section~\ref{sec:results}.

\section{Gaussian random field model}
\label{sec:gps}
Consider the spatial distribution of stagnation temperature at an isolated axial measurement plane. We denote this as $t\left(\vx\right)$, where $\vx= \left(r, \theta \right)$, with $r = \left\{r: 0\leq r \leq1 \right\}$ and $\theta = \left\{\theta: 0\leq \theta < 2\pi \right\}$ representing the non-dimensional span and circumferential location (in radians) respectively. What underpins our modelling paradigm is that the relationship between $\vx$ and $t$ has both a systematic and random component. Our overarching objective in this section is to describe the conditional distribution $p\left(t | \vx \right)$. 

We assume the existence of a set of $m$ pairwise observations of the stagnation temperature $\mathcal{D} = \left\{\left(\vx_{i}, t_{i}\right)|_{i=1}^{M} \right\}$, herewith referred to as the \emph{training data}. This set of data may also be written as $\mathcal{D}=\left(\mX, \vt \right)$ where $\mX = \left(\vx_{1}, \ldots, \vx_{M}\right)$ that is $\mX \in \mathbb{R}^{M \times 2}$, and $\vt =\left(t_1, \ldots, t_{M}\right)$ with $\vt \in \mathbb{R}^{M}$. 

Note that the distribution $t\left(\vx\right)$ cannot be observed directly as individual measurements are corrupted by a variety of noise sources as described above. Mathematically, we assume that this corruption is Gaussian, yielding $t\left(\vx\right) \sim f\left(\vx\right) + \mathcal{N}\left(0, \sigma^2\right)$ where $\sigma$ is the standard deviation associated with an individual measurement. Extending this across the training data $\mathcal{D}$ we write
\begin{equation}
    \vt \sim \left(\begin{array}{c}
f\left(\vx_{1}\right)\\
\vdots \\
f\left(\vx_{M}\right)
\end{array}\right) + \mathcal{N}\left(\mathbf{0}, \boldsymbol{\Sigma} \right)
\end{equation}
where $\boldsymbol{\Sigma} = \sigma^2 \boldsymbol{I}$, where $\boldsymbol{I} \in \mathbb{R}^{M \times M}$ is the identity matrix. In instances where noise correlations between the measurements can be inferred, $\boldsymbol{\Sigma}$ can be appropriately altered to encode such correlations, and thus need not be restricted to the identity. Given the observed non-dimensional span and circumferential locations, the likelihood of $t$ may be given as
\begin{equation}
    p\left(\vt | \vf \right) =\mathcal{N}\left( \mathbf{0}, \boldsymbol{\Sigma} \right). 
\end{equation}
We now wish to represent the conditional distribution $p\left(\vf | \mX \right)$, which is designed to capture the systematic component of the relationship between $\vx$ and $\vt$. To this end, we define a Gaussian model prior of the form
\begin{align}
    p\left(\vf | \mX \right) & = \mathcal{N} \left( \vf \left|  \left[ \begin{array}{c}
m\left(\vx_1\right) \\
\vdots\\
m\left(\vx_M\right)
\end{array}\right], \left[ \begin{array}{ccc}
k\left(\vx_1, \vx_1\right) & \cdot & k\left(\vx_{1}, \vx_{M} \right) \\
\vdots & \ddots & \vdots \\
k\left(\vx_M, \vx_1\right) & \cdot & k\left(\vx_{M}, \vx_{M} \right)
\end{array}\right] \right. \right), \\
p\left(\vf | \mX \right) & = \mathcal{N}\left( \vf | \vm, \mK \right)
\end{align}
where $m\left(\vx\right)$ and $k\left(\vx, \vx \right)$ are chosen mean and covariance functions respectively. Covariance functions are typically parameterised by certain hyperparameters, i.e., $k\left(\vx, \vx ; \boldsymbol{\psi}\right)$, values for which need to inferred based on both the data and the assumed likelihood (noise model). We define them to be $\boldsymbol{\psi}\in \mathbb{R}^{d}$ and express the model prior as $p\left(\vf | \mathbf{X}, \boldsymbol{\psi}\right)$. Note that these hyperparameters will also have a prior distribution $p\left(\boldsymbol{\psi}\right)$, which must be used to evaluate the posterior distribution $p\left(\vf | \mathcal{D}, \boldsymbol{\psi} \right)$. According to Bayes' rule, this is given by
\begin{equation}
    p\left(\vf | \mathcal{D}, \boldsymbol{\psi} \right) = \frac{p\left(\vt | \vf \right) p\left( \vf | \mX, \boldsymbol{\psi} \right) }{ p\left(\mathcal{D}, \boldsymbol{\psi}\right)}= \frac{p\left(\vt | \vf \right) p\left( \vf | \mX, \boldsymbol{\psi} \right) p\left(\boldsymbol{\psi}\right)}{ \int p\left(\vt | \vf \right) p\left( \vf | \mX, \boldsymbol{\psi} \right) p\left(\boldsymbol{\psi}\right) d\boldsymbol{\psi} \; d \vf}.
    \label{equ:bayes}
\end{equation}
where the denominator is termed the evidence or marginal likelihood; it is essentially a scaling constant. As we assumed a Gaussian likelihood, for a chosen value of $\boldsymbol{\psi}^{\ast}$, inference over $\vf$ the vector of latent values can be derived analytically
\begin{align}
    \begin{split}
    p\left(\vf | \mathcal{D}, \boldsymbol{\psi} \right) \propto \; & p\left(\vt | \vf \right) p\left( \vf | \mX, \boldsymbol{\psi}^{\ast} \right) \\
    = \; & \mathcal{N}\left(\vt | \vf, \boldsymbol{\Sigma} \right) \mathcal{N}\left( \vf | \vm, \mK \right) \\
    \propto \; & \mathcal{N}\left( \vf |\vm +  \boldsymbol{\Sigma}^{-1}\left(\mK^{-1} + \boldsymbol{\Sigma}^{-1}\right)^{-1} \left(\vt - \vm \right), \left(\mK^{-1} + \boldsymbol{\Sigma}^{-1}\right)^{-1} \right).
    \end{split}
    \label{equ:gp_derivation}
\end{align}
For simplicity, we set $\vm=0$ and zero-mean the data $\vt$. Before detailing how we can use the formulations above for predicting $\vt$ at \emph{testing} spatial locations, a few statements on equations \eqref{equ:bayes} and \eqref{equ:gp_derivation} are in order. 

Typically, when using Bayes' rule, we wish to identify the full posterior distribution $p\left(\vf | \mathcal{D}, \boldsymbol{\psi} \right)$, rather than just its moments or maximum value. To do so, one can utilise well-worn Markov chain Monte Carlo (MCMC) methods that generate samples from the prior distribution $p\left(\boldsymbol{\psi}\right)$ to inform the posterior distribution, based on a variety of factors including whether the chosen sample $\boldsymbol{\psi}^{\ast}$ yields a higher posterior density. Note that in practice $\boldsymbol{\psi}$, may also be a function of certain \emph{other} hyerperparameters, in which case priors must be assigned and duly sampled from. As a technicality, it should be noted that each value of the hyperparameters yields a Gaussian random field. From MCMC, we obtain a distribution of values for the hyperparameters, and thus a distribution of Gaussian random fields. This therefore does not yield a posterior Gaussian distribution, but a mixture of Gaussian distributions. Rather than negotiate a mixture of Gaussians, we choose to identify the single value of the hyperparameters that maximises the likelihood, given the data and the priors. This \emph{maximum a posteriori} (MAP) value yields a posterior Gaussian distribution which as we will see later results in an analytical form of the distance required for anomaly detection. It should be noted that although the MAP will likely offer a reduced estimate of the overall uncertainty, it has the advantage of delivering faster inference, which is necessary for anomaly detection. 

Once the posterior distribution or its mode has been computed, it can be used to predict the stagnation temperature at other locations. Let $\tilde{\mX} \in \mathbb{R}^{N \times 2}$ with $\tilde{\mX} = \left( \tilde{\vx}_{1}, \ldots,  \tilde{\vx}_{N} \right)^{T}$ be a set of such locations and $\tilde{\vf} \in \mathbb{R}^{N}$ the corresponding [unknown] predictive values, i.e., $\tilde{\vf} = \left(f\left(\tilde{\vx}_{1}\right), \ldots, f\left(\tilde{\vx}_{N} \right) \right)^{T}$. Evaluating the covariance function at these locations, we define $\tilde{\mK}_{ij} = k\left(\vx_{i}, \tilde{\vx}_{j} ; \boldsymbol{\psi} \right)$ and $\widetilde{\mK}_{ij}= k\left(\tilde{\vx}_{i}, \tilde{\vx}_{j}; \boldsymbol{\psi} \right)$. The predictive posterior distribution can then be written as
\begin{align}
\begin{split}
    p\left(\tilde{\vf}| \mathcal{D}, \tilde{\mX}, \boldsymbol{\psi} \right) =\; & \int p\left( \tilde{\vf} | \vf, \mX, \tilde{\mX}, \boldsymbol{\psi} \right) p\left(\vf | \mathcal{D}, \boldsymbol{\psi} \right) d\boldsymbol{\psi} \; d \vf \\
    = \; & \int \mathcal{N}\left(\left[\begin{array}{c}
\vf\\
\tilde{\vf}
\end{array}\right]\left| \; \mathbf{0},\left[\begin{array}{cc}
\mK & \tilde{\mK}\\
\tilde{\mK}^{T} & \widetilde{\mK}
\end{array}\right]\right.\right) \mathcal{N}\left(\vt | \vf, \boldsymbol{\Sigma} \right) \mathcal{N}\left( \vf | \vm, \mK \right)d\boldsymbol{\psi} \; d \vf. 
\label{equ:predictive_posterior}
\end{split}
\end{align}
Note that the first term on the right hand side in \eqref{equ:predictive_posterior} is the joint distribution of the observed $\vf$ and unobserved $\tilde{\vf}$. The covariance matrix is partitioned into four blocks to capture the covariances between the training and testing input locations. Evaluating \eqref{equ:predictive_posterior} yields
\begin{equation}
    p\left(\tilde{\vf}| \mathcal{D}, \tilde{\mX}, \boldsymbol{\psi} \right) = \mathcal{N}\left( \tilde{\vf} | \tilde{\mK}^{T}\left( \mK + \boldsymbol{\Sigma} \right)^{-1} \vt, \; \widetilde{\mK} -  \tilde{\mK}^{T}\left( \mK + \boldsymbol{\Sigma} \right)^{-1} \tilde{\mK}\right),
\end{equation}
revealing the predictive posterior mean and predictive posterior covariance.

The kernel function used in this paper is adapted from our prior work in \cite{seshadri2021bayesian} and \cite{seshadri2021bayesianmass} as it was found to capture the type of variability expected in the radial and circumferential profiles. The kernel function is expressed as a product of two kernels, one denoting the kernel in the radial direction $k_{r}\left(\vr, \vr'\right)$ and another along the circumferential direction $k_{c}\left(\boldsymbol{\theta}, \boldsymbol{\theta}'\right)$
\begin{equation}
    k\left(\vx, \vx'\right) = k_{r}\left(\vr, \vr'\right)\times k_{c}\left(\boldsymbol{\theta}, \boldsymbol{\theta}'\right).
\end{equation}

First, we introduce their circumferential kernel, which comprises a Fourier series kernel which has the form
\begin{equation}
    k_{c}\left(\boldsymbol{\theta}, \boldsymbol{\theta}'\right) = \mathbf{F}\left( \boldsymbol{\theta} \right) \boldsymbol{\Lambda}^{2} \mathbf{F}\left( \boldsymbol{\theta} \right)^{T},
\end{equation}
where 
\begin{equation}
\mathbf{F}\left(\boldsymbol{\theta}\right)=\left[\begin{array}{cccccc}
1 & sin\left(\Omega_{1}\theta\right) & cos\left(\Omega_{1}\theta\right) & \ldots & sin\left(\Omega_{k}\theta\right) & cos\left(\Omega_{k}\theta\right)\end{array}\right],
\end{equation}
with $\boldsymbol{\Omega} = \left(\Omega_{1}, \ldots,  \Omega_{k}\right)$ being the $k$ wave numbers and where $\boldsymbol{\Lambda}^{2} \in \mathbb{R}^{2K+1}$ is a diagonal matrix of hyperparameters, i.e., $\boldsymbol{\Lambda}^2= diag\left(\lambda_1^{2}, \ldots, \lambda_{2k+1}^{2} \right)$, whose values need to be determined. The squared terms are used here to show that the hyperparameters represent the variances associated with either the sine or cosine of each wave number. Note that as $\lambda^{2}_{j} \rightarrow 0$, for $j=1, \ldots, 2k+1$, implies that corresponding mode does not play an important role in the Fourier series expansion. Half normal priors are assigned for the hyperparameters $\lambda^{2}_{j} \sim \mathcal{N}^{+}\left(1\right)$; this distribution has the support $[0, \infty)$, and takes in the parameter variance as the argument.

Along the radial direction, we use the well-worn squared exponential kernel
\begin{equation}
    k_{r}\left( \vr, \vr' \right) = \sigma_{f}^2 \text{exp}\left(-\frac{1}{2l^2}\; \left(\vr - \vr' \right)^{T}\left(\vr - \vr' \right)\right)
\end{equation}
which is parameterised by two hyperparameters: a kernel noise variance $\sigma_{f}^2$ and a kernel length scale $l^2$. These are both assigned half Normal priors
\begin{equation}
    \sigma_{f}^2 \sim \mathcal{N}^{+}\left(1\right), \; \; \; \; l^2 = \mathcal{N}^{+}\left(1 \right),
\end{equation}
with variances set to 1. For convenience we set $\boldsymbol{\psi} = \left(\lambda_{1}^2, \ldots, \lambda_{2k+1}^2,\sigma_f, l \right)$ and thus the prior $p\left(\boldsymbol{\psi}\right)$ represents $2k+3$ independent half Normal distributions with a variance of 1. 

\subsection{Posterior inference via MAP}
In MAP, the objective is to solve an optimization problem over the space of hyperparameters $\boldsymbol{\psi}$ for identifying the mode of the posterior. This is done via
\begin{equation}
    \underset{\boldsymbol{\psi}}{\textrm{maximise}} \; \; \; p\left(\vt | \vf \right) p\left( \vf | \mX, \boldsymbol{\psi} \right) p\left(\boldsymbol{\psi}\right),
\end{equation}
where the terms $p\left(\vt | \vf \right) p\left( \vf | \mX, \boldsymbol{\psi} \right)$ come from \eqref{equ:bayes}. Whilst this optimisation problem is generally non-convex, its gradients may be computed---either analytically for via any automatic differentiation package---and used for accelerating the optimisation with a standard gradient-based optimiser. All results in this paper use the MAP for parameter inference.

\subsection{Bayesian area average}
One important consequence of interpreting the spatial distribution of temperature or pressure as a Gaussian random field over an annulus, is that one can derive analytical expressions for linear operators that act over the random field. This is precisely what we introduced in \cite{seshadri2021bayesian}, and describe below for convenience.

The standard area average for a thermodynamic quantity $f\left(r, \theta\right)$ at an annular axial plane is written as
\begin{align}
\begin{split}
\mathcal{A}\left(f\right) & = \frac{r_{o} - r_{i} }{\pi \left(r^2_{o} - r^2_{i} \right) } \int_{0}^{1} \int_{0}^{2\pi} f\left(r, \theta \right) v \left(r \right) dr \; d \theta \\
 & = \frac{r_{o} - r_{i} }{\pi \left(r^2_{o} - r^2_{i} \right) } \int f\left(\vx \right) v \left(r \right) d \vz
\end{split}
\label{equ:area_average}
\end{align}
where $r \in [0, 1]$, $\theta \in [0, 2\pi)$ and $v\left( r \right) = r \left(r_{o} - r_{i} \right) + r_{i}$, with $r_{i}$ and $r_{o}$ being the inner and outer radii of annular section respectively. For convenience we use the coordinates $\vx = \left(r, \theta \right)$ as before. One can re-write \eqref{equ:area_average} as a linear operator acting upon the spatially varying quantity. Now, recall the joint distribution \eqref{equ:predictive_posterior} based on an available training dataset $\mathcal{D}$. We can apply the same linear operator across the posterior predictive distribution to arrive at
\begin{equation}
p\left(\mathcal{A}\left(f \right) | \mathcal{D}, \boldsymbol{\psi} \right) =\mathcal{N}\left( \underbrace{\tilde{\vw}^{T}\left( \mK + \boldsymbol{\Sigma} \right)^{-1} \vt}_{\mu_{\mathcal{A}\left(p \right) }}, \;  \underbrace{\omega -  \tilde{\vw}^{T}\left( \mK + \boldsymbol{\Sigma} \right)^{-1} \tilde{\vw} }_{\sigma^2_{\mathcal{A}\left(p \right)}} \right),
\label{equ:area_avg}
\end{equation}
where
\begin{align}
\begin{split}
    \tilde{\vw} & = \frac{r_{o} - r_{i} }{\pi \left(r^2_{o} - r^2_{i} \right) } \int k\left(\vx, \vz \right) v \left(r \right) d \vz, \; \; \; \text{where} \; \; \; \tilde{\vw} \in \mathbb{R}^{M \times 1}, \\ 
    \omega & = \left( \frac{r_{o} - r_{i} }{\pi \left(r^2_{o} - r^2_{i} \right) } \right)^2 \int \int k\left(\vz, \vz' \right) v^2 \left(r \right) d \vz \; d \vz', \; \; \; \text{where} \; \; \; \omega \in \mathbb{R}.
\end{split}
\end{align}
To clarify, for a given value of the hyperparameters, the Bayesian area average is a Gaussian distribution where the mean and variance can be calculated by plugging in the values of hyperparameters $\boldsymbol{\psi}$ into \eqref{equ:area_average}.

\section{The Wasserstein distance and optimal transport}
\label{sec:opt_transport}
In this section we introduce the proposed methodology for spatial anomaly detection. The overarching idea is based on the distance between two probability distributions, one which represents a baseline while the other represents an observed sample. We can think of the baseline akin to a \emph{gold standard}, as it represents an idealised distribution of a quantity---i.e., what we \emph{expect} the quantity to be. In what follows we define the chosen distance metric and detail our data-driven strategy for classification.


Optimal transport is the study of moving a collection of related items from one configuration into another. This movement may entail items from one discrete distribution to another discrete distribution; from one discrete distribution to one continuous distribution, or from one continuous distribution to another continuous one. These collections may include images, graphs, or more generally probability distributions---both discrete or continuous. Optimal transport has recently seen applications in myriad of diverse fields including signal processing, statistical machine learning, computer vision, and medical sciences \cite{kolouri2017optimal}. It is extremely useful for comparing signals across different coordinate systems and signal families, and thus has naturally seen some application in anomaly detection \cite{schlegl2019f}. Our exposition below closely follows the notation in \cite{peyre2019computational}.

To offer a deeper understanding of optimal transport, consider two measures $\alpha$ and $\beta$. A measure can be either a discrete or a continuous distribution and it need not integrate to unity. That said, the measure should be integrable against any continuous function and yield a real-valued output as the integral. We assume that the measures $\alpha$ and $\beta$ are defined over spaces $\mathcal{X}$ and $\mathcal{V}$ respectively. We further assume that these spaces are equipped with a distance metric. Mathematically, we state that over the set of Radon measures $\mathcal{R}\left(\mathcal{X}\right)$ and $\mathcal{R}\left(\mathcal{V}\right)$ we have $\alpha \in \mathcal{R}\left(\mathcal{X}\right)$ and $\beta \in \mathcal{R}\left(\mathcal{V}\right)$. For a given $\mathcal{X}$, let   $\mathcal{R}\left(\mathcal{X}\right)_{+}$ denotes the space of all the positive measures, while $\mathcal{R}\left(\mathcal{X}\right)_{+}^{1}$ denotes the space all positive measures that satisfy $\int_{\mathcal{X}} d \alpha = 1$ for all $\alpha \in \mathcal{R}\left(\mathcal{X}\right)_{+}^{1}$ \cite{peyre2019computational}. 

\begin{figure}
\begin{center}
\includegraphics[scale=0.50]{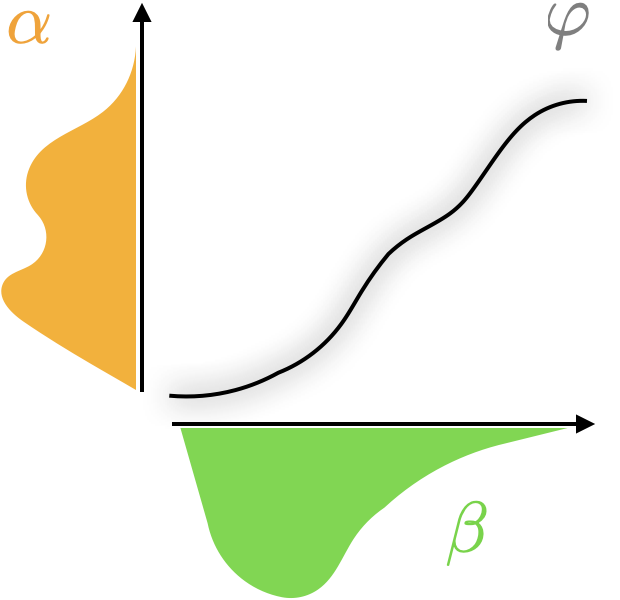}
\caption{A schematic of optimal transport for two continuous distributions.}
\label{fig:ot}
\end{center}
\end{figure}

There are two key ideas in optimal transport: conservation of mass when transporting elements of $\alpha$ to $\beta$, and the ability to split mass when doing so---also termed the Kantorovich relaxation which is particularly suited for discrete measures. Thus, we seek a transport map $T$ that pushes all the mass of $\alpha$ towards the mass $\beta$, through which the mass itself may be split. To crystallise the relationship over the two measures, we consider couplings $\varphi \in \mathcal{R}_{+}^{1}\left( \mathcal{X} \times \mathcal{V} \right)$ which represent all the joint distributions over the product space of the marginals $\mathcal{X} \times \mathcal{V}$. The notation $\varphi \in \mathcal{U}\left[\alpha, \beta \right]$, encodes the mass conservation constraint, i.e., $\varphi$ is uniformly distributed between $\alpha$ and $\beta$ (see Figure~\ref{fig:ot}). 

For random samples $\vx \in \alpha$ and $\vv \in \beta$, the optimal transport problem is
\begin{equation}
   \mathcal{L}\left(\alpha, \beta \right) := \underset{\varphi \; \in \; \mathcal{U}\left[\alpha, \beta\right]}{\text{minimum}}\; \left\{ \int_{\mathcal{X} \times \mathcal{Y}} c\left(\vx, \vv \right) \; d \varphi\left(\vx, \vv \right) \right\},
\end{equation}
as the minimisation of a distance metric subject to a certain cost function $c\left(\vx, \vv \right)$. If we consider the standard $L_{\rho}$-norm distance between $\alpha$ and $\beta$, then the optimal minimiser, should it exist, is given by the $L_{\rho}$-th Wasserstein distance
\begin{equation}
    W_{L_{\rho}}\left( \alpha, \beta\right) = \underset{\varphi \; \in \; \mathcal{U}\left[\alpha, \beta\right]}{\text{minimum}}\; \left\{ \int_{\mathcal{X} \times \mathcal{Y}}  \left\Vert \vx - \vv \right\Vert_{L_{\rho}}^{\rho} \; d \varphi\left(\vx, \vv \right) \right\}^{\frac{1}{\rho}}
\end{equation}

\subsection{Optimal transport with multivariate Gaussians}
\label{sec:opt_mv_gauss}
\begin{figure}
\begin{center}
\includegraphics[scale=0.35]{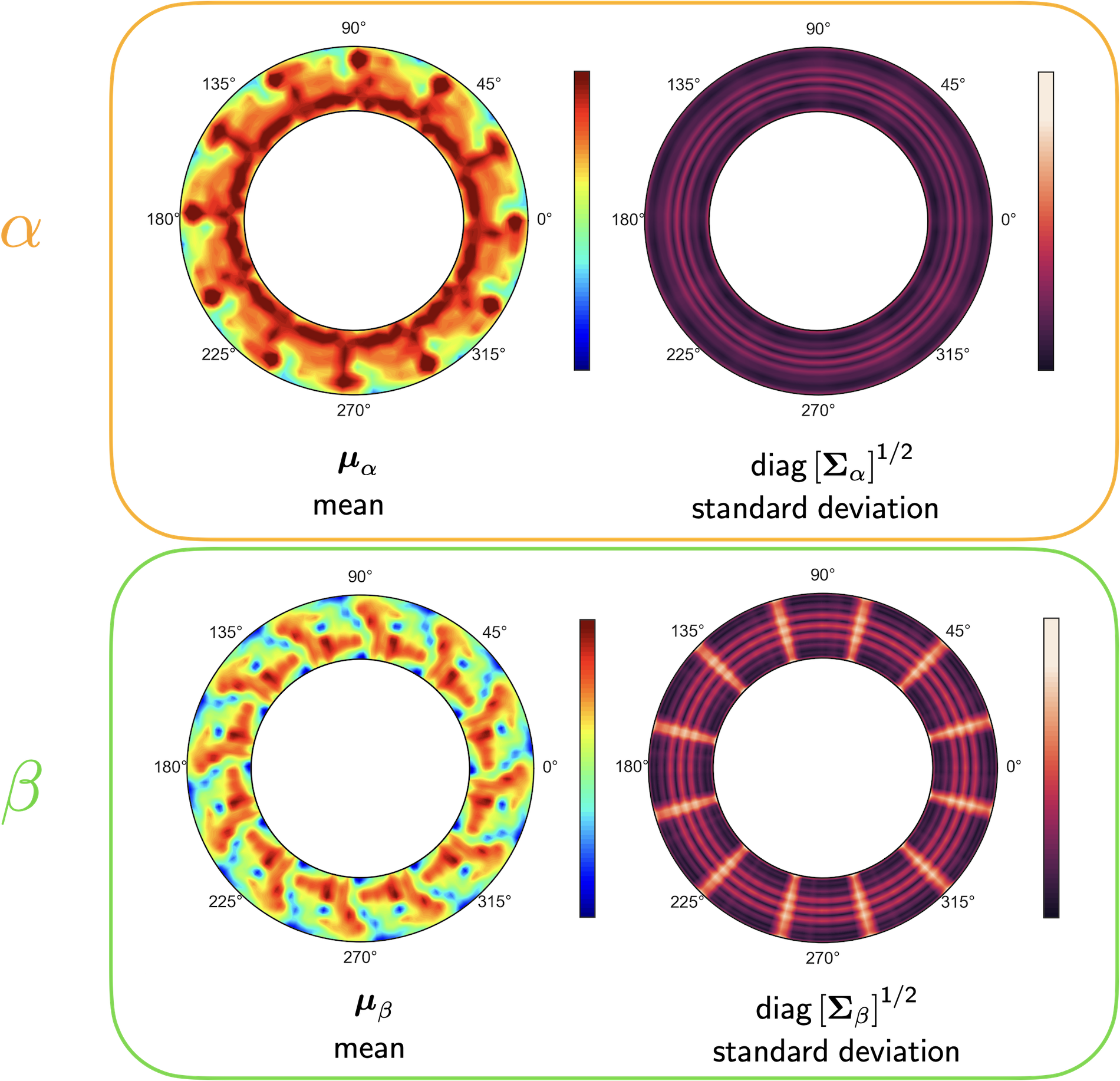}
\caption{A schematic of two Gaussian annular random fields.}
\label{fig:ot2}
\end{center}
\end{figure}

A closed form expression for the Wasserstein metric exists when evaluating the distance between two Gaussian distributions with $L_{\rho}=2$. Let us define two Gaussian annular random fields (see Figure~\ref{fig:ot2}) for a thermodynamic quantity $\alpha = \mathcal{N} \left( \boldsymbol{\mu}_{\alpha}, \boldsymbol{\Sigma}_{\alpha}
 \right)$ and $\beta = \mathcal{N} \left( \boldsymbol{\mu}_{\beta}, \boldsymbol{\Sigma}_{\beta}\right)$, where $\boldsymbol{\mu}_{\alpha} \in \mathbb{R}^{N}, \boldsymbol{\mu}_{\beta} \in \mathbb{R}^{N}, \boldsymbol{\Sigma}_{\alpha} \in \mathbb{R}^{N \times N}$ and $\boldsymbol{\Sigma}_{\beta} \in \mathbb{R}^{N \times N}$. We also assume that both covariance matrices $\boldsymbol{\Sigma}_{\alpha}$ and $\boldsymbol{\Sigma}_{\beta}$ are symmetric positive definite. By construction, the spaces $\mathcal{X}$ and $\mathcal{Y}$ are equivalent. The Wasserstein distance (see page 34 in \cite{peyre2019computational}) between them using a quadratic cost, is given by
\begin{equation}
W_{2}^2\left(\alpha, \beta \right) := \left\Vert \boldsymbol{\mu}_{\alpha} - \boldsymbol{\mu}_{\beta}\right\Vert _{2}^{2} + \text{tr}\left( \boldsymbol{\Sigma}_{\alpha} \right) + \text{tr}\left( \boldsymbol{\Sigma}_{\beta} \right) - 2\text{tr}\left( \boldsymbol{\Sigma}_{\alpha}^{1/2} \boldsymbol{\Sigma}_{\beta} \boldsymbol{\Sigma}_{\alpha}^{1/2}\right)^{1/2},
\label{equ:wasserstein_distance}
\end{equation}
where the superscript $1/2$ denotes the matrix square root, and the expression $\left\Vert \cdot \right\Vert_{2}^{2}$ denotes the sum of the squares of the argument $\left( \cdot \right)$. Note that this is equivalent to the Bures-Wasserstein distance between covariance matrices $\boldsymbol{\Sigma}_{\alpha}$ and $\boldsymbol{\Sigma}_{\beta}$.

\begin{figure}
\begin{center}
\includegraphics[scale=0.45]{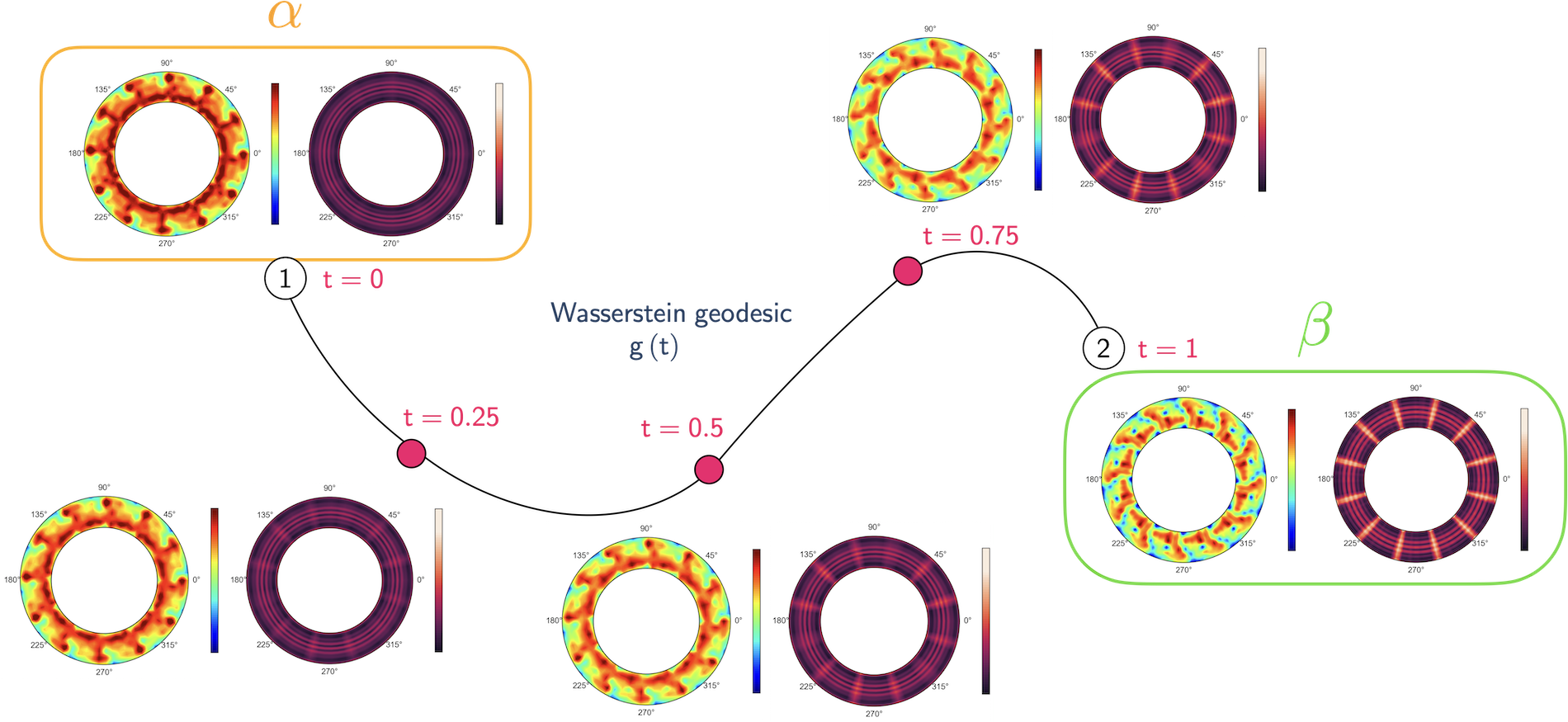}
\caption{A schematic of showing the temporal map between two distributions $\alpha$ and $\beta$.}
\label{fig:ot3}
\end{center}
\end{figure}

One can interpret the \emph{movement} of probability mass as a pseudo-temporal map $g$ connecting $\alpha$ and $\beta$, where $g\left(0\right) = \alpha$ and $g\left(1\right) = \beta$. The map $g$  is parameterised by a scalar \emph{time} parameter $\left\{t: 0 \leq t \leq 1 \right\}$ for which $g\left(t \right)$ returns the probability mass at time $t$. The resulting path from $t=0$ to $t=1$ is called the Wasserstein geodesic, as shown in Figure~\ref{fig:ot3}. For a sample $\zeta_{0}$ from the distribution $\alpha$, the temporal movement is given by
\begin{equation}
\zeta_{t} = \left(1 - t \right)\zeta_{0} + t \mH,
\label{equ:transport}
\end{equation}
where $\zeta_{t}$ is the \emph{transported} sample, and $\mH$ is the optimal transport map. For two Gaussian distributions, the optimal transport map is given by
\begin{align}
\mH = \boldsymbol{\mu}_{\beta} + \mR  \left( \zeta_{0} - \boldsymbol{\mu}_{\alpha} \right), \; \; \; \text{with} \; \; \; 
\mR  = \boldsymbol{\Sigma}_{1}^{-1/2} \left( \boldsymbol{\Sigma}_{\alpha}^{1/2}  \boldsymbol{\Sigma}_{\beta}  \boldsymbol{\Sigma}_{\alpha}^{1/2} \right)^{1/2} \boldsymbol{\Sigma}_{\alpha}^{-1/2}.
\end{align}
As the displacement of each sample in \eqref{equ:transport} is affine, i.e., 
\begin{equation}
\zeta_{t} = \zeta_{0} \left( 1 - t + t\mR \right) + \left(t \boldsymbol{\mu}_{\beta} - t \mR \boldsymbol{\mu}_{\alpha} \right),
\end{equation}
the distribution for any $t$ is Gaussian with moments given by
\begin{equation}
p\left(t\right)  = \mathcal{N} \left( \left( 1 - t \right)\boldsymbol{\mu}_{\alpha} +  t \boldsymbol{\mu}_{\beta}, \; \left( \left(1 - t\right)\mI + t\mR \right)^{T} \boldsymbol{\Sigma}_{\alpha} \left( \left(1 - t\right)\mI  + t\mR \right) \right),
\label{equ:temporal_dist}
\end{equation}
where $\mI \in \mathbb{R}^{N \times N}$ is the identity matrix. It is straightforward to show that when $t=0$ the right hand side of \eqref{equ:temporal_dist} is $\mathcal{N}\left( \boldsymbol{\mu}_{\alpha}, \boldsymbol{\Sigma}_{\alpha} \right)$ and setting $t=1$ yields $\mathcal{N}\left( \boldsymbol{\mu}_{\beta}, \boldsymbol{\Sigma}_{\beta} \right)$. 

\subsection{Fusing multiple distributions via a weighted barycenter}
\label{sec:barycenter}
As comparisons between distributions are inherently done in pairs, it will be useful to ensure that the baseline measurement is a good representation of possibly distinct, yet completely non-anomalous measurements. We utilise ideas within optimal transport as a means to fuse multiple \emph{gold standard} measurements into a single representative one.

The task of computing a representative distribution from a set of distributions is analogous to the idea of computing the centroid via k-means clustering for data (see Remark 9.2 in \cite{peyre2019computational}). For a collection of samples the mean or the \emph{barycenter} is the minima of a weighted sum of the distances between a candidate and all the samples. To clarify, let $\alpha_1, \ldots, \alpha_{k}$ denote the set of $K$ input distributions for which we wish to compute the barycenter. We define the weighted barycenter $\alpha_{\star}$ as
\begin{align}
    \underset{\alpha_{\star} \; \in \; \mathcal{M}_{+}^{1} \left( \mathcal{X} \right) }{\text{minimise}} \; \sum_{j=1}^{K}\vartheta_{j} \mathcal{L}\left( \alpha_{\star}, \alpha_{j} \right),
    \label{equ:barycenter_defn}
\end{align}
with weights $\vartheta_1, \ldots, \vartheta_{M}$, where $\vartheta_{j} \geq 0$ and $\sum_{j=1}^{K} \vartheta_{j} = 1$. When the weights in \eqref{equ:barycenter_defn} are equal, we say that $\alpha_{star}$ is the barycentre of $\alpha_1, \ldots, \alpha_{K}$. Additionally, note that in the case where $\mathcal{X} = \mathbb{R}$ and $L_{\rho}=2$, then under certain circumstances the barycenter is unique \cite{agueh2011barycenters}. 

\subsection{Computing the barycenter with multivariate Gaussians}
As shown in Agueh and Carlier \cite{agueh2011barycenters}, for a collection of Gaussian distributions the Wasserstein barycenter is Gaussian $\mathcal{N}\left(\boldsymbol{\mu}_{\star}, \boldsymbol{\Sigma}_{\star} \right)$ with known mean $\boldsymbol{\mu}_{\star}$ and covariance $\boldsymbol{\Sigma}_{\star}$. Consider a set of $K$ Gaussian distributions $\alpha_{j}= \mathcal{N}\left(\boldsymbol{\mu}_{j}, \boldsymbol{\Sigma}_{j} \right)$ for $j=1, \ldots, K$. The mean and covariance of the barycenter is then given by
\begin{equation}
\boldsymbol{\mu}_{\star} = \sum_{j=1}^{K} \vartheta_{j} \boldsymbol{\mu}_{j}, \;\; \; \; \boldsymbol{\Sigma}_{\star} = \underset{\boldsymbol{\Sigma}}{\text{minimise}}\; \sum_{j=1}^{K} \vartheta_{j} \text{tr} \left(\boldsymbol{\Sigma} + \boldsymbol{\Sigma}_{j} - 2\left[ \boldsymbol{\Sigma}^{1/2} \boldsymbol{\Sigma}_{j} \boldsymbol{\Sigma}^{1/2} \right]^{1/2} \right).
\label{equ:barycenter_mean_cov}
\end{equation}
The weights $\vartheta_{j} \geq 0$ can be set uniform, i.e., $\vartheta_{j} = 1/K$ or they may be chosen to appropriately weight certain distributions; we discuss this salient point in the context of generating synthetic samples later. Whilst the mean in \eqref{equ:barycenter_mean_cov} is trivially computed, the covariance requires some clarification. Whilst a closed-form analytical solution is not available, an iterative solution is at hand. Following the fixed point technique in \cite{alvarez2016fixed}, we compute $\boldsymbol{\Sigma}_{\star} $ through iterative updates via
\begin{equation}
\boldsymbol{\Sigma}_{\star}^{(i + 1)} = \sum_{j=1}^{K} \vartheta_{j} \left( \left( \boldsymbol{\Sigma}_{\star}^{(i)}\right)^{1/2} \boldsymbol{\Sigma}_{j}  \left( \boldsymbol{\Sigma}_{\star}^{(i)}\right)^{1/2}\right)^{1/2},
\label{equ:bary_cov_iterative}
\end{equation}
where the superscript $(i)$ denotes the present iterate. Whilst no convergence proof for \eqref{equ:barycenter_mean_cov} exists, convergence is obtained in practice \cite{alvarez2016fixed}.

It may be useful to also compute the Bayesian area average of the barycenter, which is also a univariate Gaussian. Let 
\begin{equation}
   \mathcal{N}\left( \mathcal{\mu}_{\mathcal{A}\left(\alpha_i\right)}, \;  \sigma^2_{\mathcal{A}\left(\alpha_i\right)}\right), \; \; \; \text{for} \; \; \; i=1, \ldots, K
\end{equation}
denote the Bayesian area average for the $K$ distributions above. We can then write the barycenter's area average as
\begin{equation}
    \mathcal{N}\left( \sum_{i=1}^{K}\vartheta_{i}\mathcal{\mu}_{\mathcal{A}\left(\alpha_i\right)}, \;  \left(\sum_{i=1}^{K} \left(\vartheta_{i}\sigma^2_{\mathcal{A}\left(\alpha_i\right)}\right)^{1/2}\right)^2 \right);
\end{equation}
here the variance does not require a fixed point iteration and is analytically solved for.


\section{Spatial anomaly detection via optimal transport}
\label{sec:anomaly}
In this section we make precise our framework for spatial anomaly detection. Following our introductory remarks, this requires a distance metric and a threshold. 

\subsection{Distance metric for anomaly detection}
We begin by formalising the anomaly detection problem. Given $S$ sensors with spatial locations $\mathring{\mX} = \left( \mathring{\vx}_{1}, \ldots, \mathring{\vx}_{S} \right)^{T}$, we wish to ascertain which sensor(s) is anomalous based on its location and thermodynamic readings $\mathring{\vf} = \left( \mathring{f}_{1}, \ldots, \mathring{f}_{S} \right)^{T}$. To do so, we use the Bayesian inference framework in \ref{sec:gps} to arrive at an observed Gaussian random field $\beta$. We assume that we have also have access to the Gaussian random field arising from some baseline measurements which we hold to be representative of what one would expect. We term this Gaussian random field $\alpha$. Let the computed area average means for the observed and baseline Gaussian random fields be given by $\mu_{\mathcal{A}\left(\beta\right)}$ and $\mu_{\mathcal{A}\left(\alpha\right)}$ (computed via \eqref{equ:area_average}). The distance metric we propose is a quintessentially a weighted 1D Wasserstein distance evaluated at the same spatial locations $\left(\vr, \boldsymbol{\theta}\right)$ across $\alpha$ and $\beta$ of the form $\vd = \left(d_1, \ldots, d_{S}\right)$, where
\begin{equation}
    d_j = \left(\frac{\boldsymbol{\mu}_{\alpha}\left( \mathring{\vx}_{j}\right) }{\mu_{\mathcal{A}\left(\alpha\right)}}- \frac{\boldsymbol{\mu}_{\beta}\left( \mathring{\vx}_{j} \right) }{\mu_{\mathcal{A}\left(\beta\right)}}\right)^2 + \frac{\boldsymbol{\Sigma}_{\alpha}\left(\mathring{\vx}_{j}, \mathring{\vx}_{j} \right)  }{\mu^{2}_{\mathcal{A}\left(\alpha\right)}} +\frac{\boldsymbol{\Sigma}_{\beta}\left(\mathring{\vx}_{j}, \mathring{\vx}_{j} \right)  }{\mu^{2}_{\mathcal{A}\left(\beta\right)}} - \frac{2 \left(\boldsymbol{\Sigma}_{\alpha}\left(\mathring{\vx}_{j}, \mathring{\vx}_{j} \right) \cdot \boldsymbol{\Sigma}_{\beta}\left(\mathring{\vx}_{j}, \mathring{\vx}_{j} \right)\right)^{1/2}}{\mu_{\mathcal{A}\left(\alpha\right)}\mu_{\mathcal{A}\left(\beta\right)}}
    \label{equ:dist_metric}
\end{equation}
for $j=1, \ldots, S$. There are a few remarks to make regarding the proposed distance metric. First, this metric is a 1D analogue of the Wasserstein distance presented in \eqref{equ:wasserstein_distance}, and by construction provides a distance between the distributions of $\alpha$ and $\beta$ indexed by each sensor's location. Note that the measurements used to infer $\beta$ and $\alpha$ can be distinct, as they are based solely on their respective predictive posteriors at $\mathring{\mX}$. Second, to mitigate the relatively large penalty imposed by subtracting the square of the means in \eqref{equ:wasserstein_distance}, we normalise all the mean terms by the corresponding area average mean and normalise all the variance terms by the square of the corresponding area average mean. This in practice should facilitate comparisons between engines that have slightly different means without flagging them as anomalous.


\subsection{Setting the threshold}
Once $\vd$ has been computed for an observed set of data, a delineation has to be made with regards to whether any of the $S$ sensors are yielding anomalous values. This paper adopts a relatively straightforward data-driven approach to set the threshold. Given a repository of $Q$ data sets, $\mathcal{D} = \left\{\mathcal{D}_{1}, \ldots, \mathcal{D}_{Q} \right\}$, all of the form previously shown, we can evaluate $\vd$ in \eqref{equ:dist_metric} for each pair, yielding $\left\{\vd_1, \ldots, \vd_{ Q \choose 2} \right\}$\footnote{The notation $Q \choose 2$ denotes the number of combinations of $Q$ data sets in pairs without repetitions, i.e., $Q! / \left(2! (Q - 2)! \right)$.} values. We assume that a majority of these $Q$ datasets are standard, however, there are a few anomalous ones included too. Then we proceed to calculate the $95\%$ percentile value of all the aggregated distances, and use that as our threshold value $\tau$. In other words, if $d_i \geq \tau$ then it is likely anomalous.

\subsection{Synthetic data generation via manifold sampling}
\begin{figure}
\begin{center}
\includegraphics[scale=0.38]{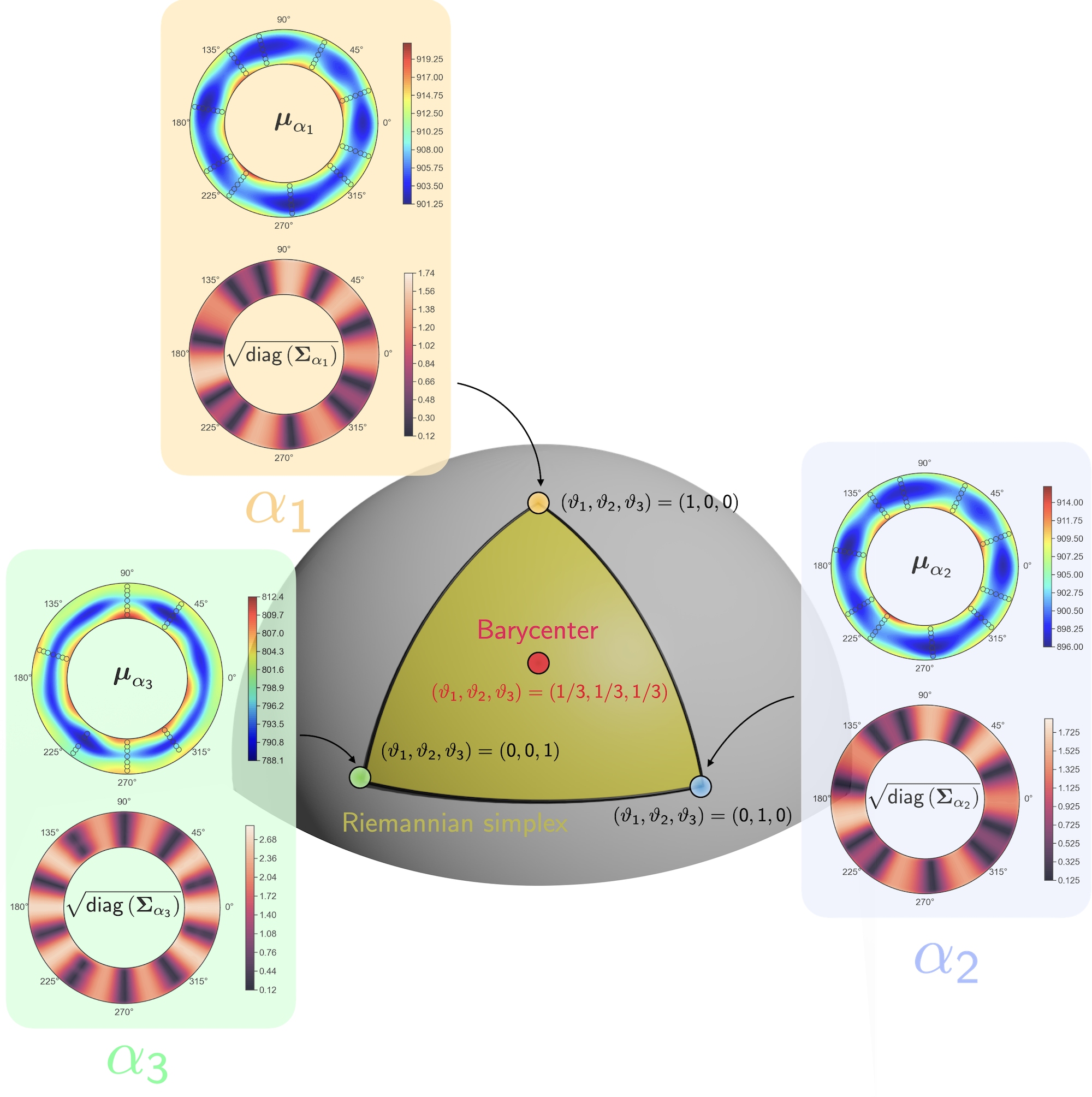}
\caption{A schematic of a Riemannian simplex (shaded yellow region) with $K=3$ distributions $\alpha_1, \alpha_2, \alpha_3$. Each vertex represents a Gaussian annular random field where the mean and standard deviations are shown across the annulus. The grey hemisphere is shown to contrast a standard Euclidean simplex with the Riemannian one. Each edge of the Riemannian simplex denotes the geodesic between any two vertices.}
\label{fig:ot4}
\end{center}
\end{figure}

We envisage a shortage of quality training data, and thus offer a recipe to synthetically generate more data. Recall in section \ref{sec:opt_mv_gauss}, the Wasserstein geodesic was introduced as an affine transformation between two distributions $\alpha$ and $\beta$; parameterised by a pseudo-temporal parameter $t$. In section \ref{sec:barycenter}, formulas for computing the barycenter associated with multiple distributions was provided. One can generalise these ideas to a Riemannian simplex \cite{dyer2015riemannian}, which comprises $K$ vertices given by distributions $\alpha_1, \ldots, \alpha_{K}$ and the inner \emph{geodesic convex} hull, i.e., the \emph{space} bounded by the simplex edges. To generate samples, we assign $\vartheta_{1}, \ldots \vartheta_{K}$ to be Dirchlet distributed random variables, and solve \eqref{equ:barycenter_mean_cov}. The resulting distributions are guaranteed to fill the Riemannian simplex (see Figure~\ref{fig:ot4} for a schematic).

\section{Demonstration on engine data}
\label{sec:results}
Following a brief overview of the measurements used, we present results of our spatial anomaly detection framework on real engine data. We also illustrate how one can generate synthetic data using the ideas above.
\subsection{Measurement and data overview}
\label{sec:measurements}
All the data shown here corresponds to steady-state temperature measurements taken at a given engine thrust level. The greater the thrust, the higher the temperatures at each station. We define a single engine test as being a run up (and often down) the power curve, and therefore comprising numerous engine extracts. An extract represents data collected when the engine is effectively adiabatic, i.e., it has had time to stabilise at the required operating thrust. 

At a given extract, the measurements are obtained by sampling all the thermocouple voltages at 192 kHz with a rolling average running for the last 20 milliseconds. After filtering the signal to remove noise and electrical system artefacts, it is averaged over a 30 second interval at a rate of 33 Hz. Subsequently, these millivolt values are converted into Kelvin via a series of calibrations that cover static, batch-wire and recovery effects. This yields both a mean stagnation temperature for each measurement and an uncertainty. For the same thrust level, at a given measurement station, the uncertainties for all sensors are assumed to be similar.

Our dataset is born from stagnation temperature measurements taken from a multitude of engines that had been allocated as development test assets, in which a high level of gas path instrumentation had been provisioned to gain insight into the engine functional behaviour. As a given engine design will have a limited number of these test assets, due to their significant cost, we make use of similarity of the various recent projects to increase our sample size. Here we denote each of these different projects as A, B, C and D. Each engine project will contain numerous of these physical test assets, run over multiple builds where some of the hardware or instrumentation is swapped out to achieve a particular test aim in each build. Each test asset build will likely undergo numerous tests for which data is collected. While the basic architecture of each test asset build, including the number of rotating and stationary components in each sub-system, are the same, there will be variations in the geometry of the constituent components beyond those implied by the manufacturing tolerances to ensure the specific test objectives of each build are achieved. It is also a fact that the exact flow conditions at which the measurements were taken on one test extract will be almost impossible to match on a subsequent test, due to the geometry changes between assets, but also other factors such as the ambient conditions. This makes the task of anomaly detection more challenging, as minor variations are permitted. As we treat each data set independently—regardless of whether they are two tests from the same build or two tests from distinct builds—we do not distinguish between test number or build in what follows.

In terms of the measurements themselves, we concern ourselves with high power measurements as this represents airplane cruise conditions. For the purposes of our example, we focus on stations within the engine that were provisioned with gas path instrumentation across several projects and several test assets to provide us with data to train for a threshold and then test this on data not used in the training phase. In this paper we study three different axial stations, aptly named station 1, station 2 and station 3.

\subsection{Extracting thresholds from engine A}
\begin{figure}
\begin{center}
\begin{subfigmatrix}{2}
\subfigure[]{\includegraphics[width=.45\textwidth]{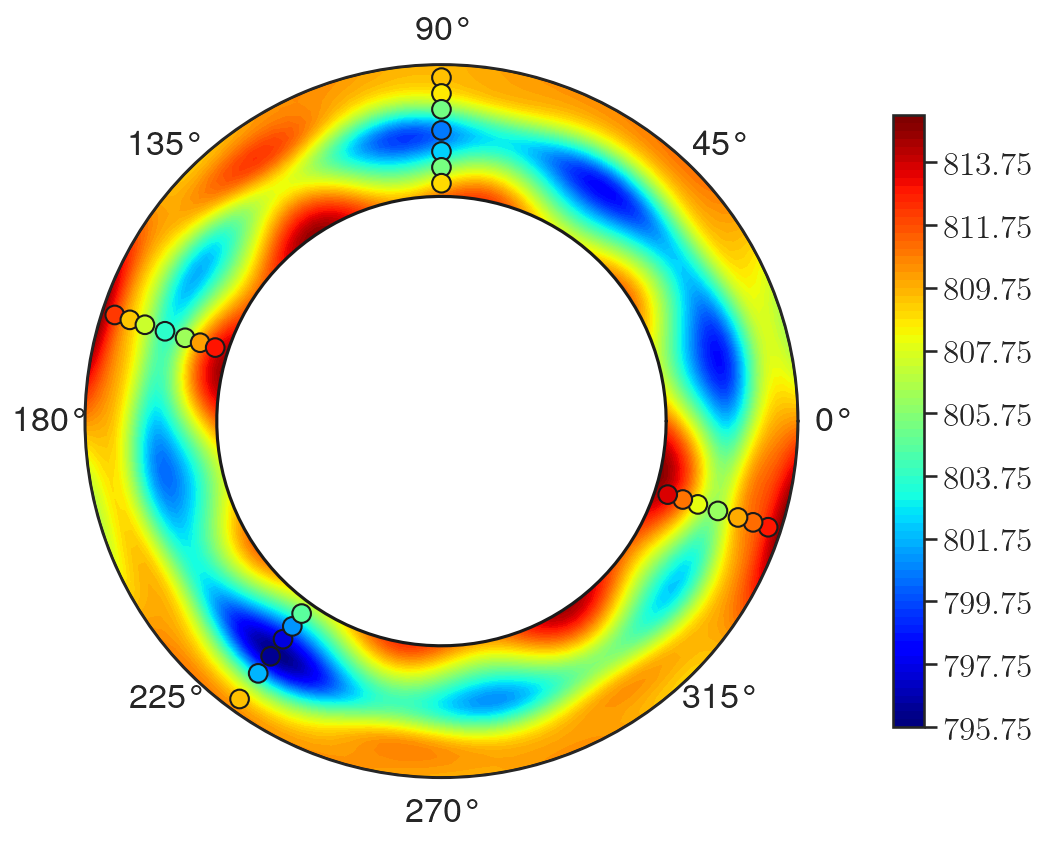}}
\subfigure[]{\includegraphics[width=.45\textwidth]{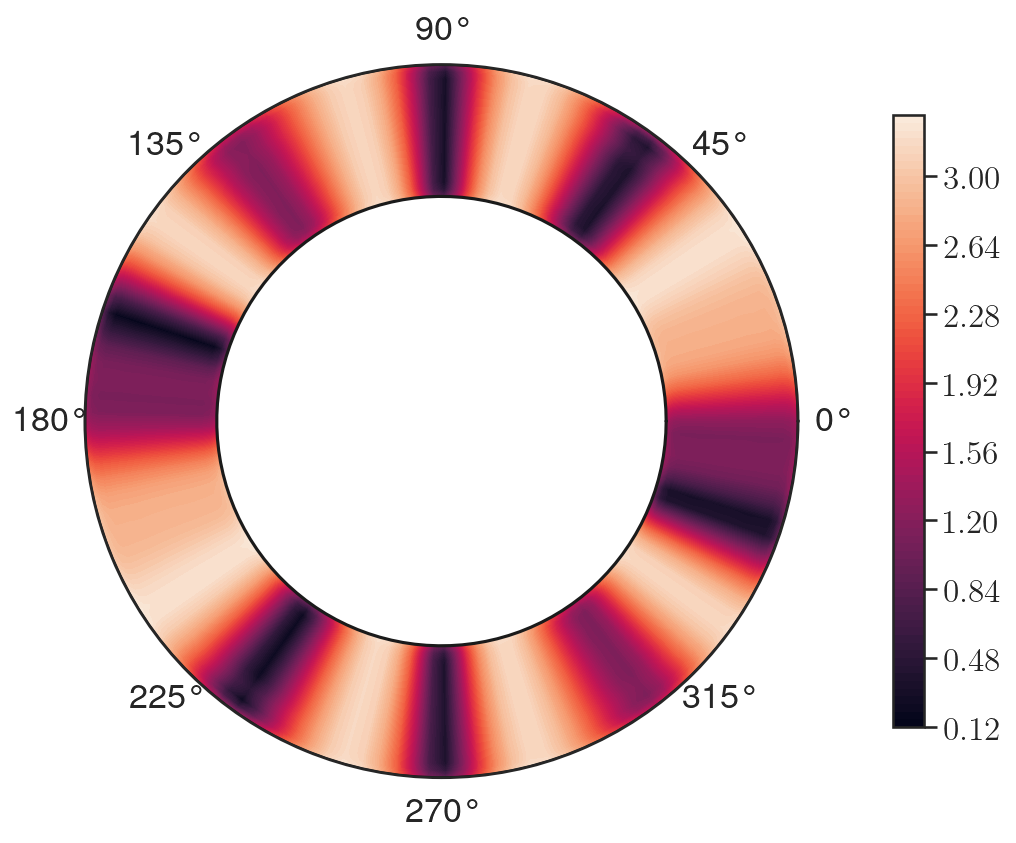}}
\subfigure[]{\includegraphics[width=.45\textwidth]{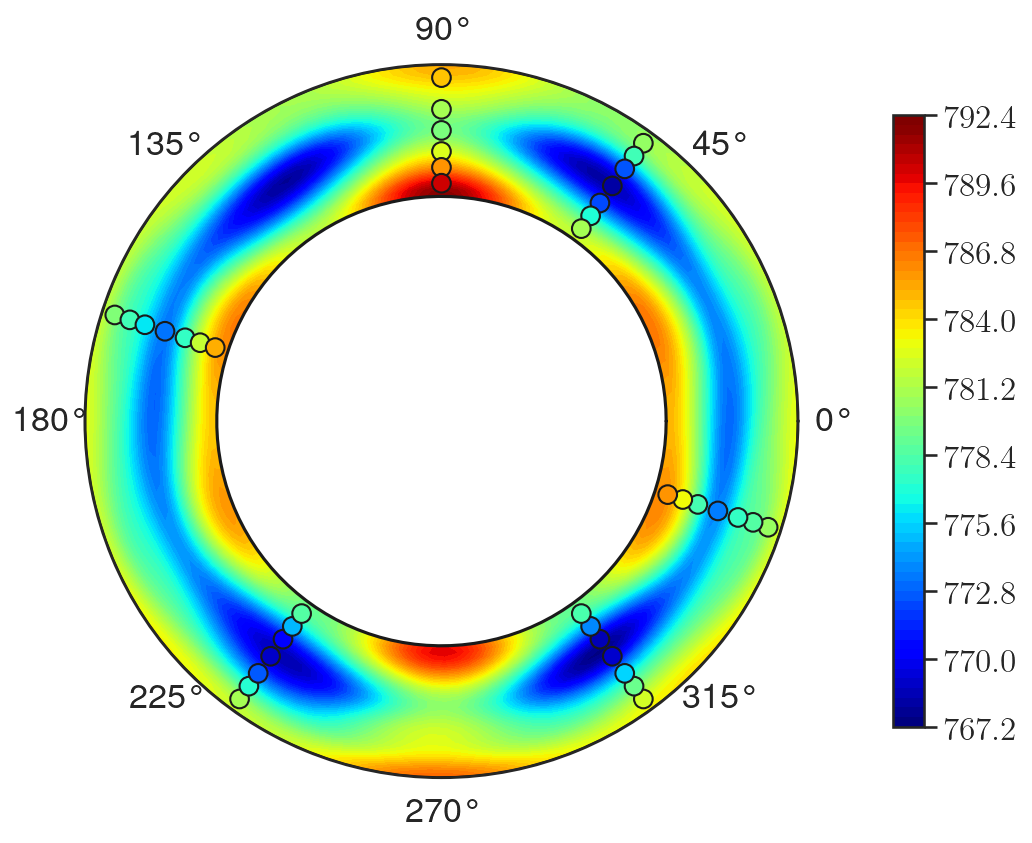}}
\subfigure[]{\includegraphics[width=.45\textwidth]{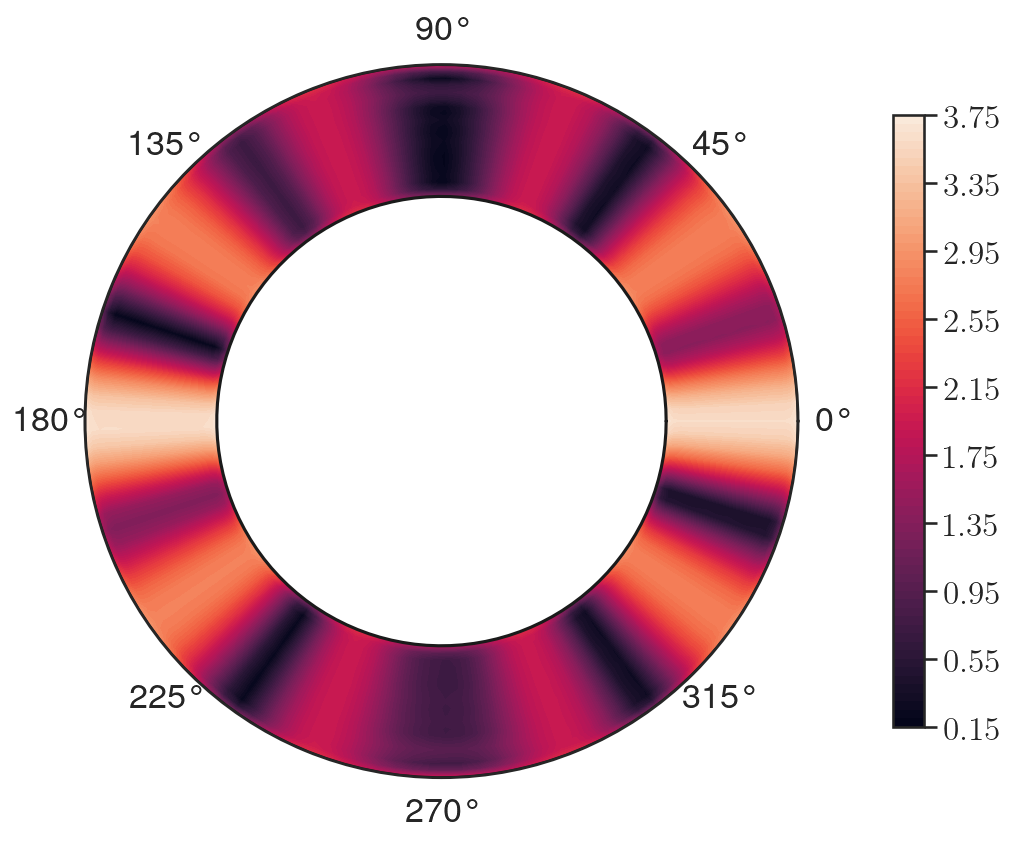}}
\subfigure[]{\includegraphics[width=.45\textwidth]{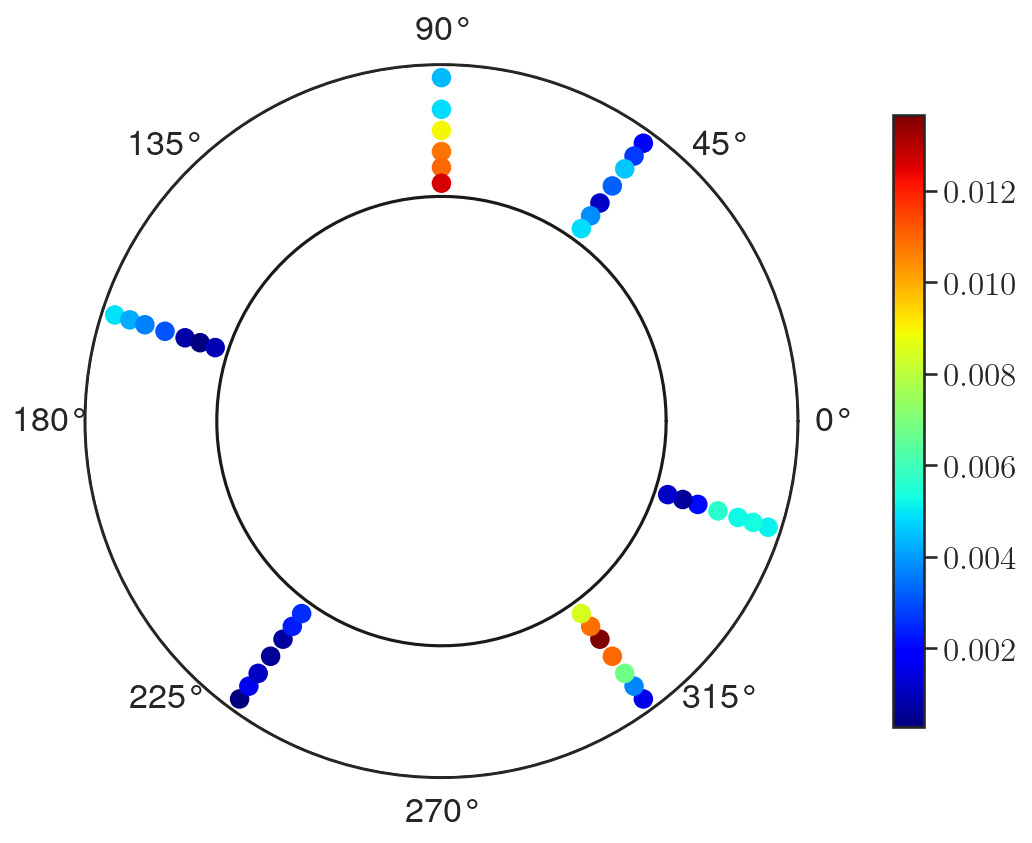}}
\end{subfigmatrix}
\caption{Computing $\vd$ for a pair of builds from engine product A station 1: (a) mean of build 348; (b) standard deviation of build 348; (c) mean of build 565; (d) standard deviation of build 565; (e) distances between the two builds $\vd$ evaluated at the sensor locations of the second build.}
\label{fig:exposition_station_1}
\end{center}
\end{figure}

As a majority of the data we have available is from engine project A, we train exclusively on it and test on the engines projects B, C, D and E. To begin, consider the subfigures in Figure~\ref{fig:exposition_station_1} that chart our training workflow. Subfigure (a) shows the radial and circumferentially placed sensors, i.e., $\mathcal{D} = \left\{\left(\vx_{i}, t_{i}\right)|_{i=1}^{M} \right\}$ with $M=27$ across the annulus as coloured markers for the data from a test asset build (termed build 348) at station 1, where build 348 is one of the assets for engine project A. The interpolated spatial field here represents the posterior predictive mean of the Gaussian random field. The interpolated spatial field here represents the posterior predictive mean of the Gaussian random field. Subfigure (b) shows the posterior predictive standard deviation; both the mean and standard deviations are evaluated using \eqref{equ:predictive_posterior}. For these results, we set $\boldsymbol{\Omega}= \left(1, 2, 3, 4, 5, 6, 7, 8\right)$ and $\sigma^2=0.04$. Similar plots are shown in subfigure (c) and (d) for a second test asset build (termed build 565) at the same instrumentation location in the engine (station 1), which is also for an engine project A test asset using the same values of $\boldsymbol{\Omega}$ and $\sigma^2$. In subfigure (e) the values of $\vd$ are shown across the different spatial locations associated with the sensor positions in (c). It should be clear that these distance values represent a continuous spectrum of possible distances that we wish to threshold via an appropriately chosen scalar parameter $\tau$, i.e., if a given distance $d_i \geq \tau$ for any $i=1, \ldots, S$ then it is flagged as anomalous. 

\begin{figure}
\begin{center}
\begin{subfigmatrix}{2}
\subfigure[]{\includegraphics[width=.45\textwidth]{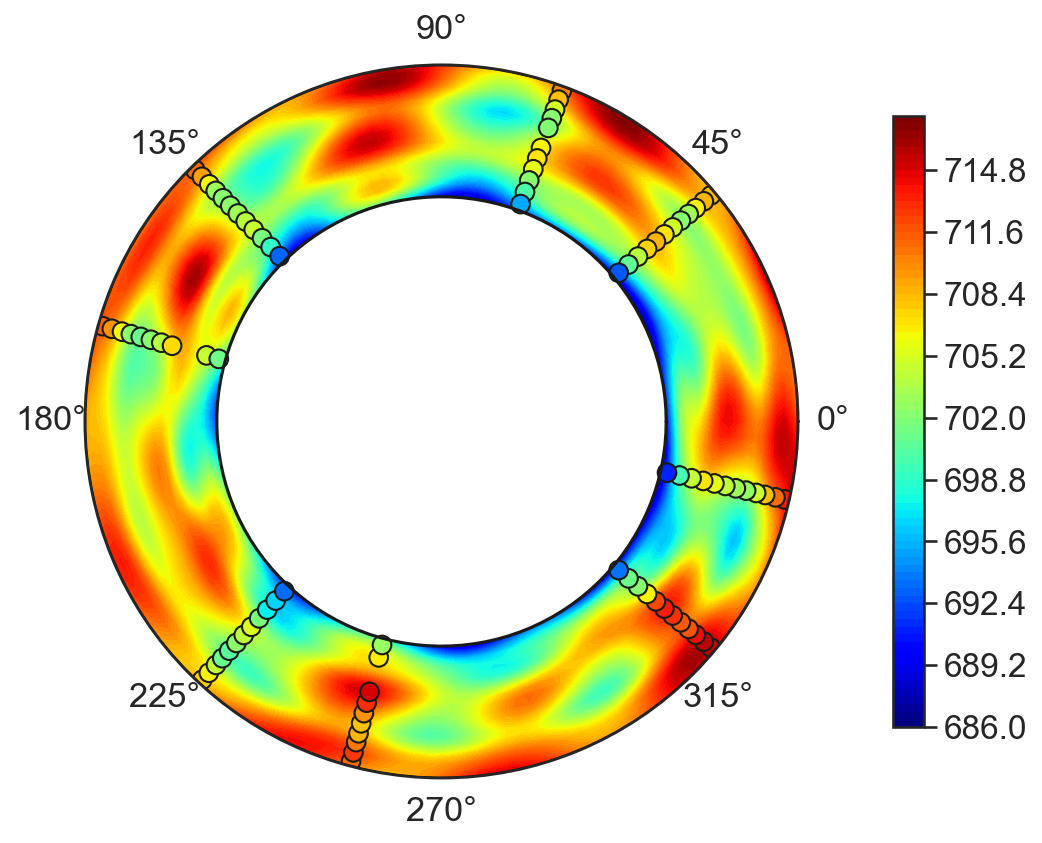}}
\subfigure[]{\includegraphics[width=.45\textwidth]{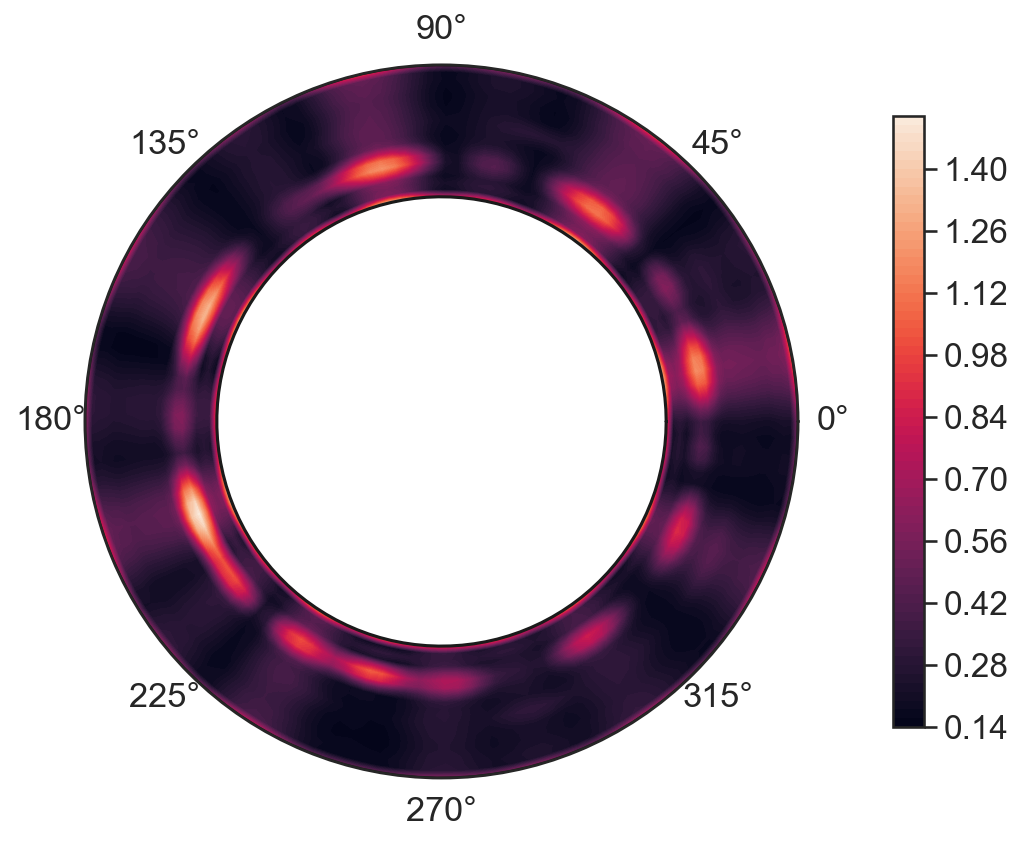}}
\subfigure[]{\includegraphics[width=.45\textwidth]{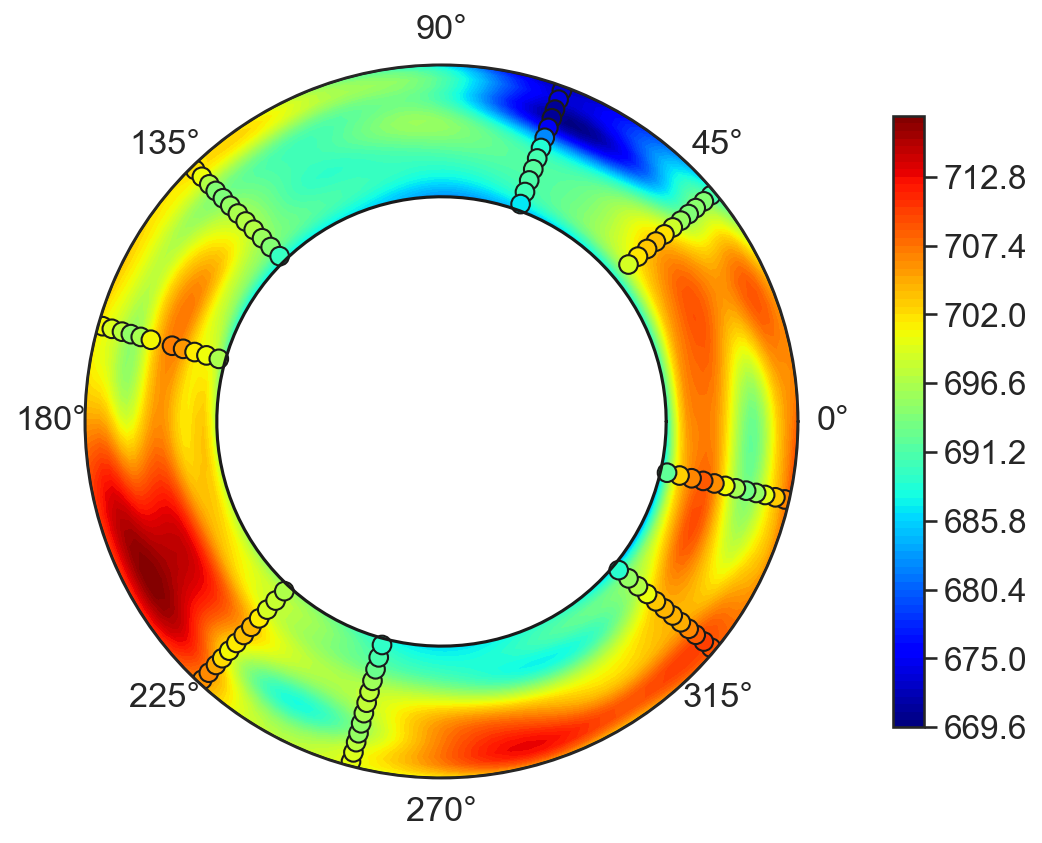}}
\subfigure[]{\includegraphics[width=.45\textwidth]{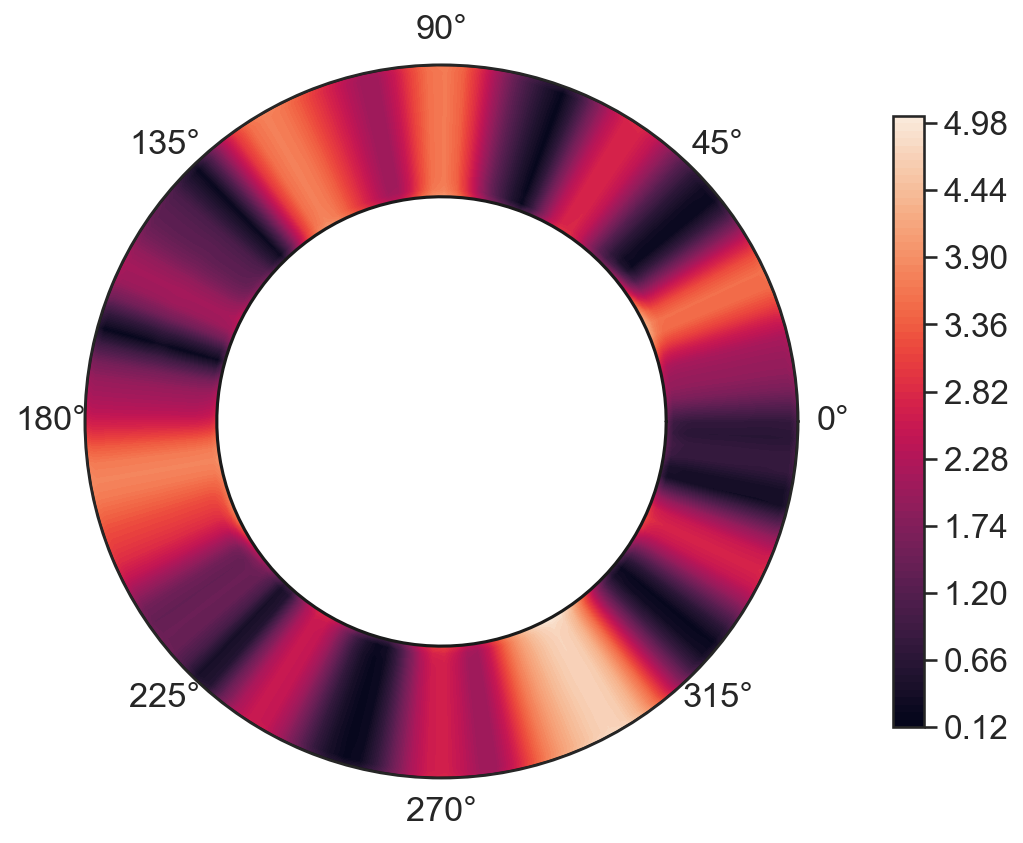}}
\subfigure[]{\includegraphics[width=.45\textwidth]{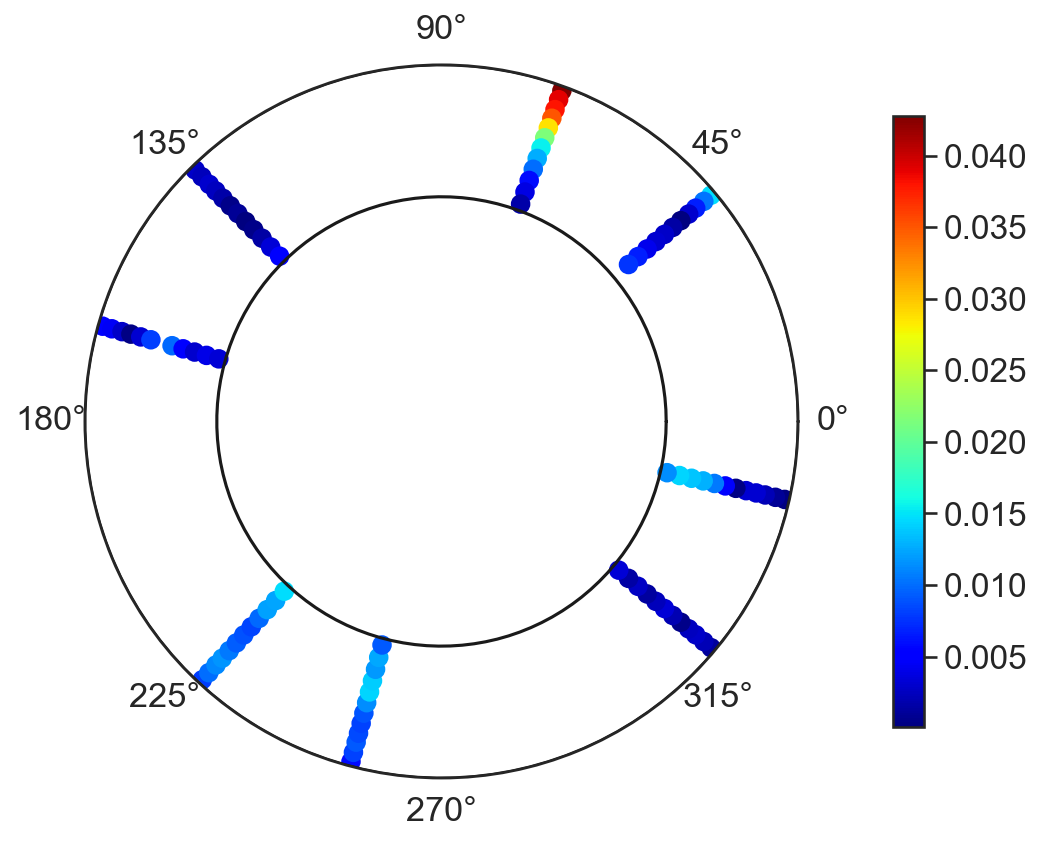}}
\end{subfigmatrix}
\caption{Computing $\vd$ for a pair of builds from engine A station 3: (a) mean of build 145; (b) standard deviation of build 145; (c) mean of build 565; (b) standard deviation of build 565; (d) distances between the two builds $\vd$ evaluated at the sensor locations of the second build.}
\label{fig:exposition_station_3}
\end{center}
\end{figure}

A similar workflow at station 3 for another pair of measurements from two test assets from engine project A is shown in Figure~\ref{fig:exposition_station_3}. Here we set $\boldsymbol{\Omega}= \left(1, 2, 3, 4, 5, 6, 7, 8, 9, 10, 11\right)$ and $\sigma^2=0.035$. One observation we make is that across different engine stations, the distance values are distinct in magnitude warranting a bespoke $\tau$ parameter for each station. The wave numbers and noise for station 2 is set to be the same as those set for station 3.

\begin{figure}
\begin{center}
\begin{subfigmatrix}{2}
\subfigure[]{\includegraphics[]{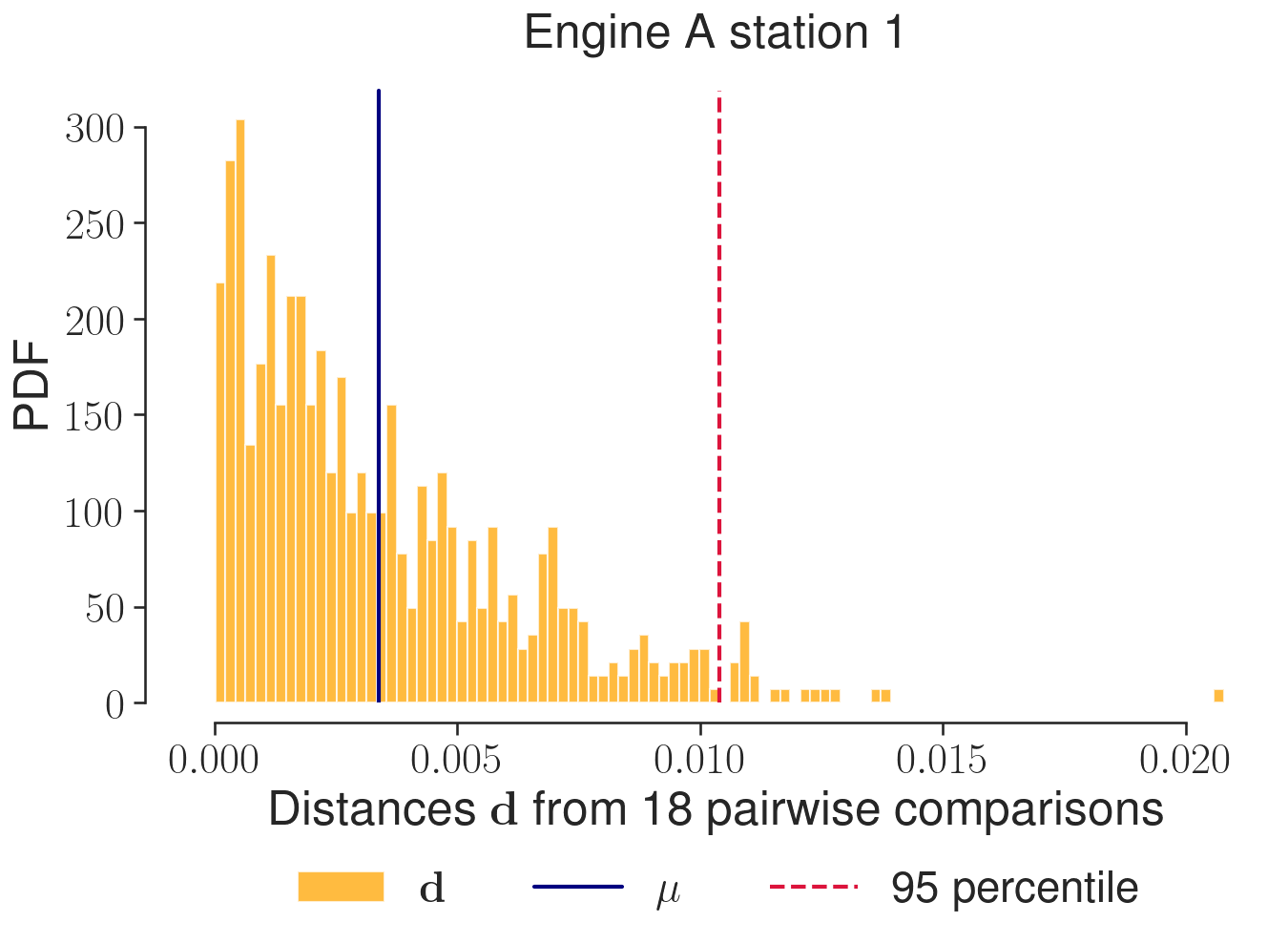}}
\subfigure[]{\includegraphics[]{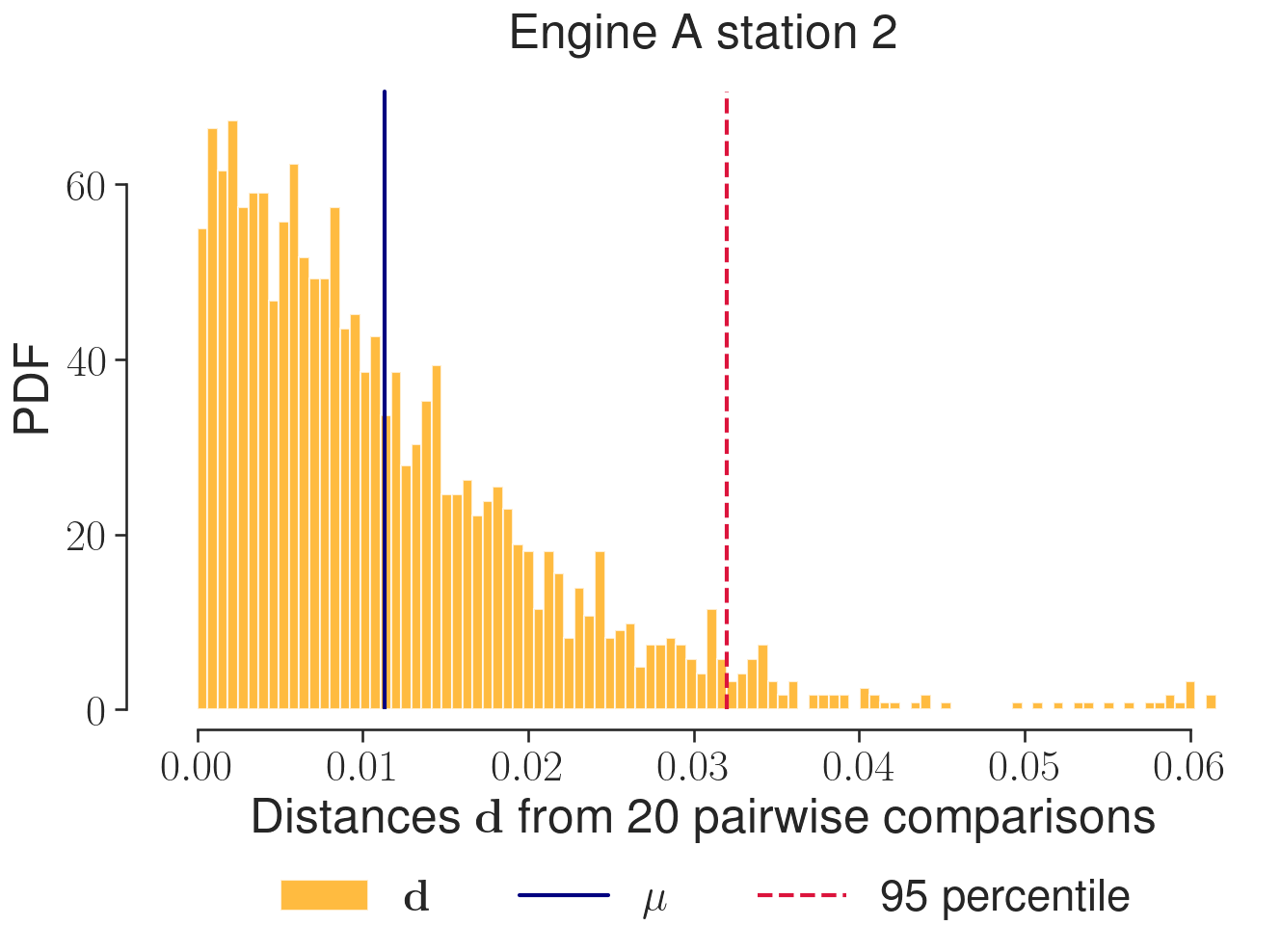}}
\subfigure[]{\includegraphics[]{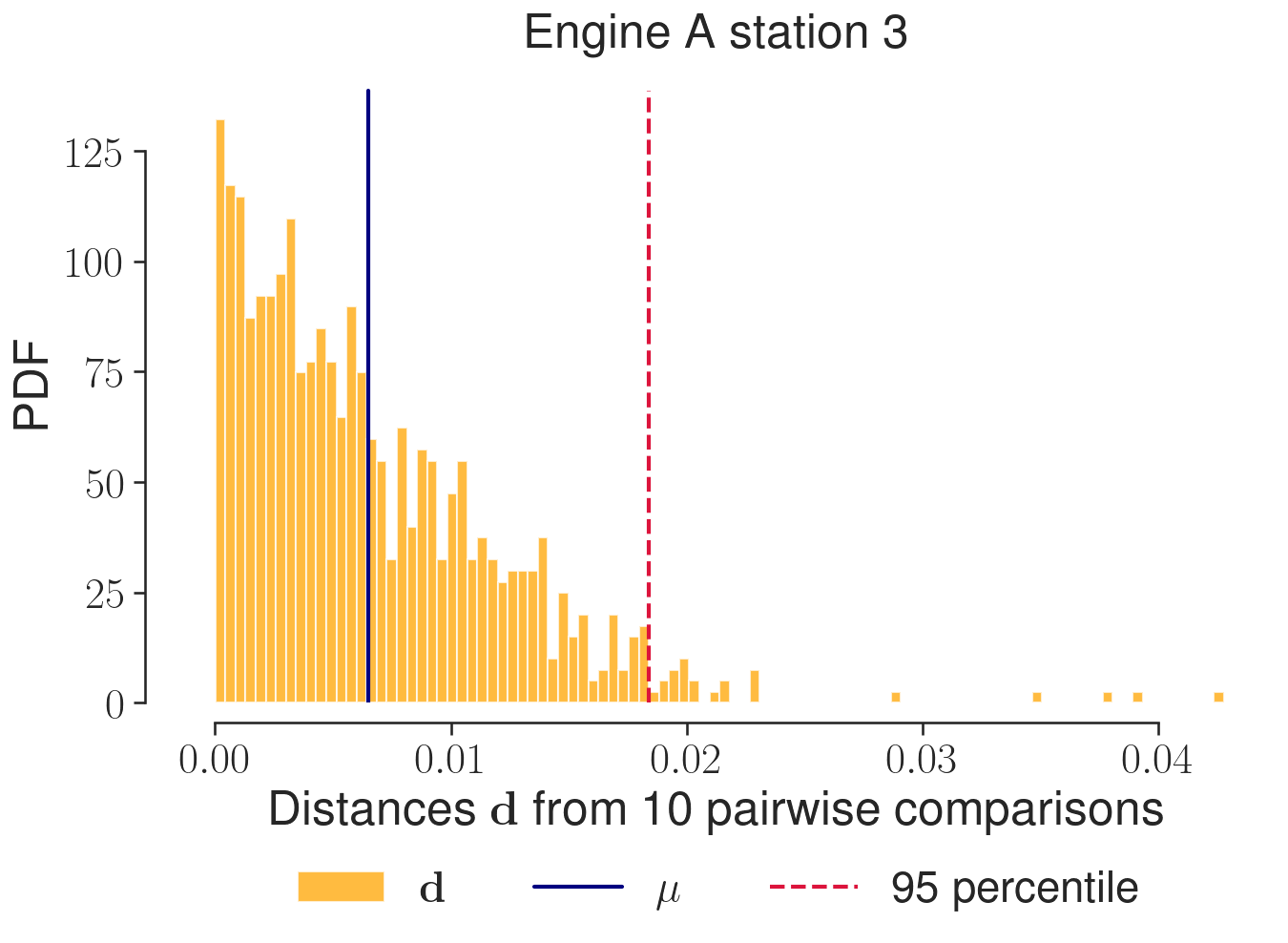}}
\end{subfigmatrix}
\caption{Aggregated pairwise distance values $\vd$ across numerous engine data sets for engine stations: (a) 1; (b) 2; (c) 3. The number of pairwise comparisons vary as not every engine test will have instrumentation at a given station.}
\label{fig:distance_histogram}
\end{center}
\end{figure}

We aggregate the distance values obtained from numerous pairwise comparisons for the three axial stations, and plot them as histograms in Figure~\ref{fig:distance_histogram}. The $95\%$ percentile value associated with each of the three stations is also shown. For station 1, we set $\tau=0.0103$, at station 2 $\tau=0.032$, and for station 3 $\tau=0.0184$.

\subsection{Demonstrating anomaly detection on engines B, C, D and E}
Here we demonstrate the utility of our approach on a few test cases---all on different engines families. While distinct, there are similar characteristics across these engines which make them suitable candidates to test the framework, even though the different values for $\tau$ were ascertained solely from engine A.

\begin{figure}
\begin{center}
\begin{subfigmatrix}{2}
\subfigure[]{\includegraphics[width=.45\textwidth]{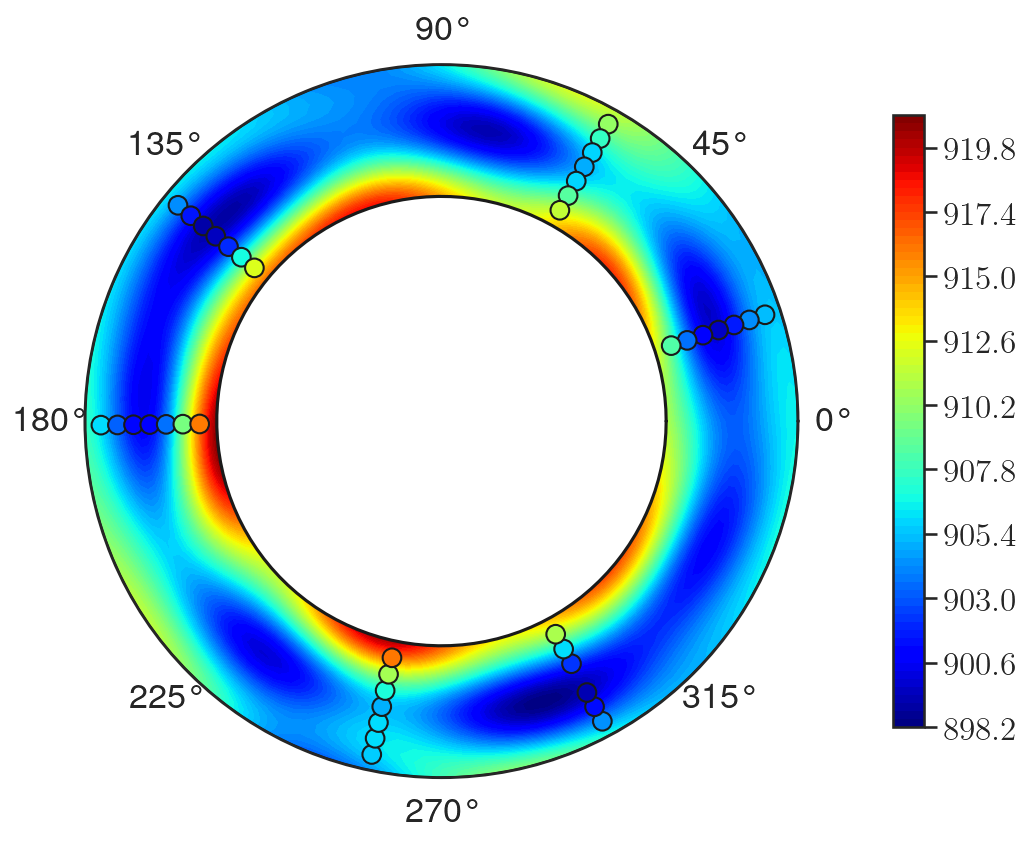}}
\subfigure[]{\includegraphics[width=.45\textwidth]{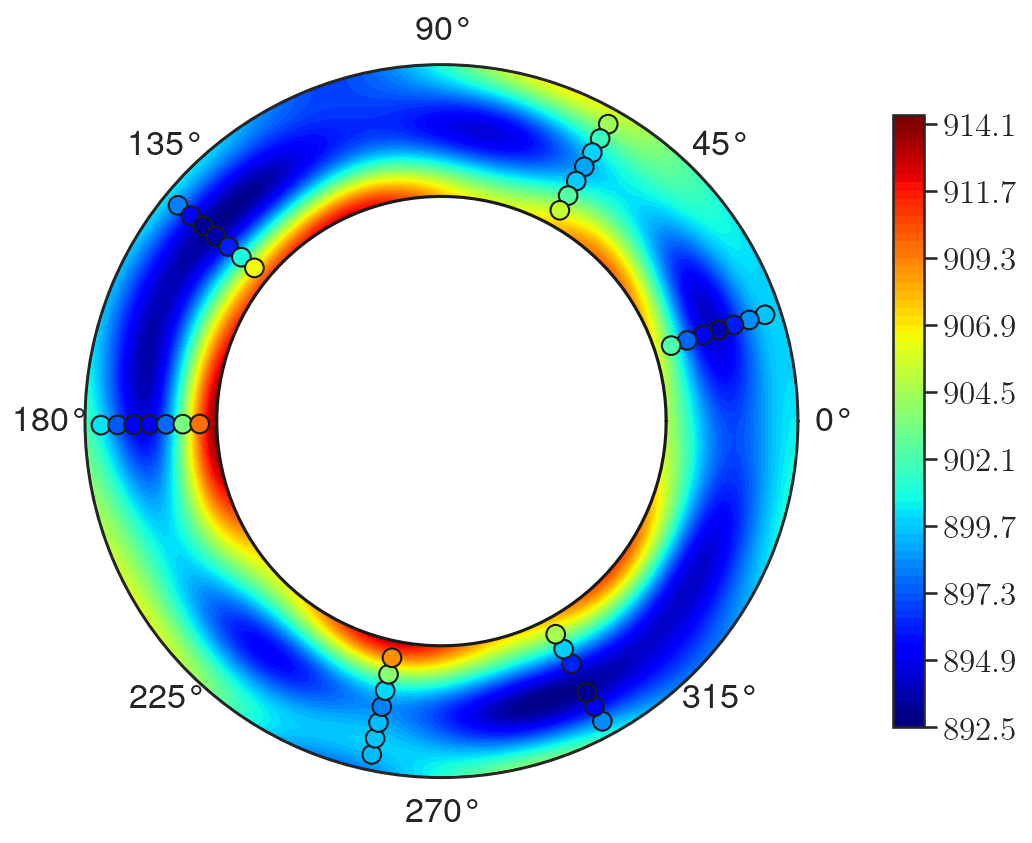}}
\subfigure[]{\includegraphics[width=.46\textwidth]{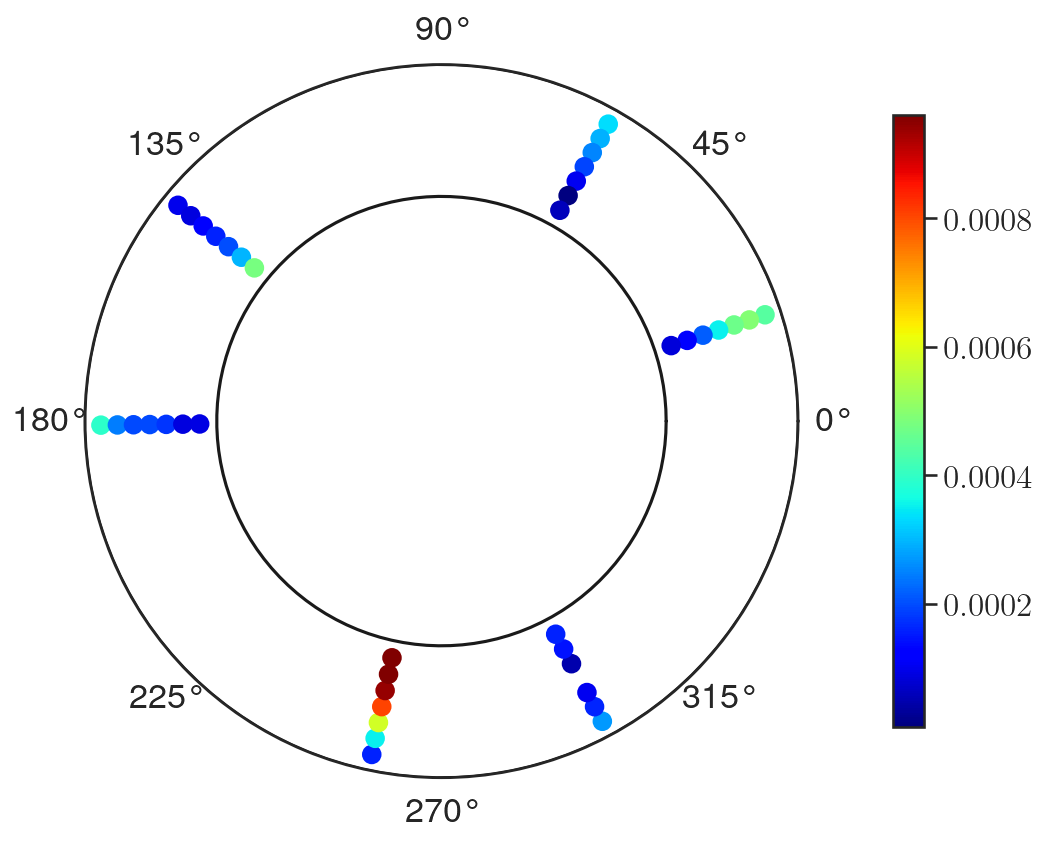}}
\subfigure[]{\includegraphics[width=.46\textwidth]{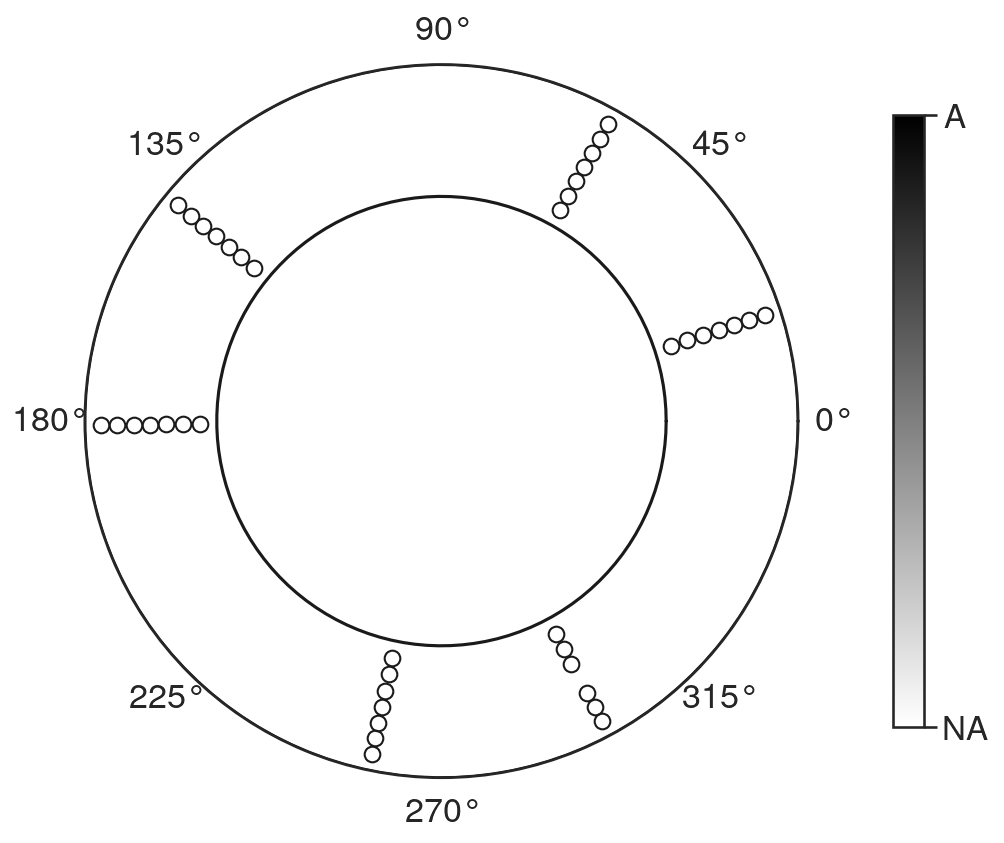}}
\end{subfigmatrix}
\caption{Testing on engine B station 1: (a) mean of build 283; (b) mean of build 278; (c) distances between the two builds $\vd$ evaluated at the sensor locations of the second build; (d) classified anomalies based on $\tau$ (NA: not anomalous; A: anomalous).}
\label{fig:exposition_station_A}
\end{center}
\end{figure}

We consider two very similar test asset builds (termed build 283 and 278) in Figure~\ref{fig:exposition_station_A} for station 1 from engine project B to demonstrate that the threshold chosen is sufficient for not yielding false positives, i.e., it is not overly penalising. Figures~\ref{fig:exposition_station_A}(a) and (b) show the posterior predictive means for two very similar measurements, with the computed values of $\vd$. As none of the distance values exceed the threshold of $\tau=0.103$ for this station, none of the sensor positions in Figure~\ref{fig:exposition_station_A}(d) are classified as anomalous (A), but are classified as not anomalous (NA). Note that this is a binary classification, and the apparent colour gradient in this subfigure should be ignored.

\begin{figure}
\begin{center}
\begin{subfigmatrix}{2}
\subfigure[]{\includegraphics[]{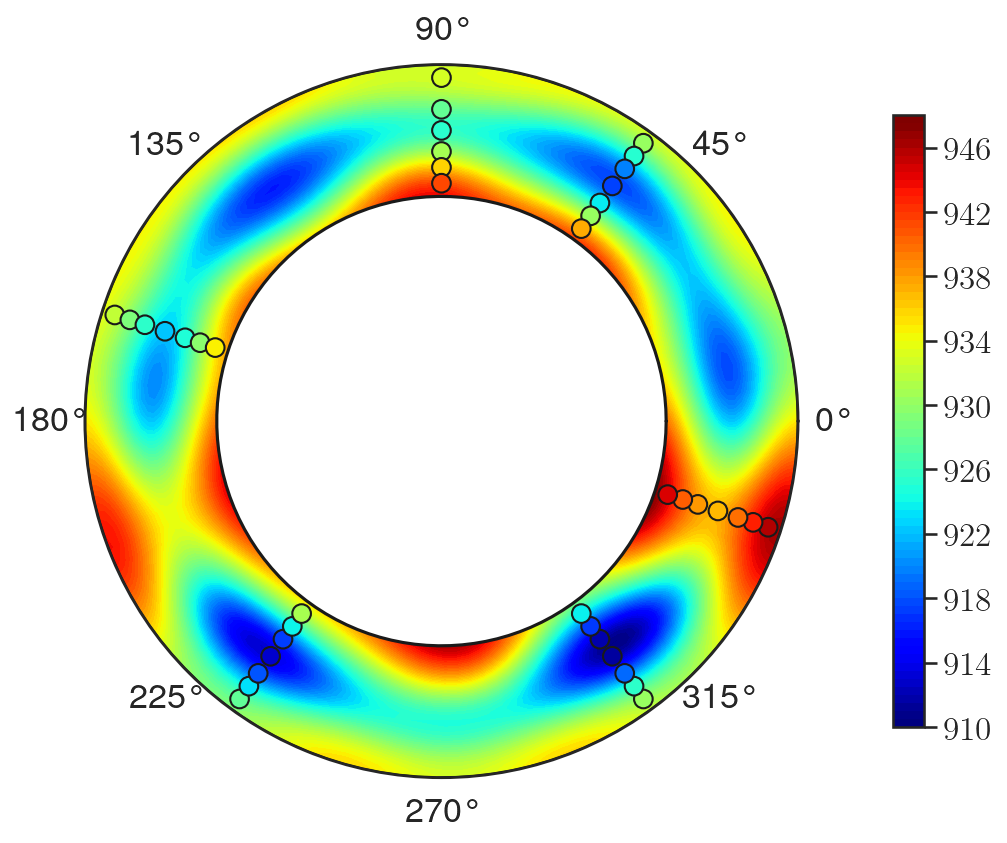}}
\subfigure[]{\includegraphics[]{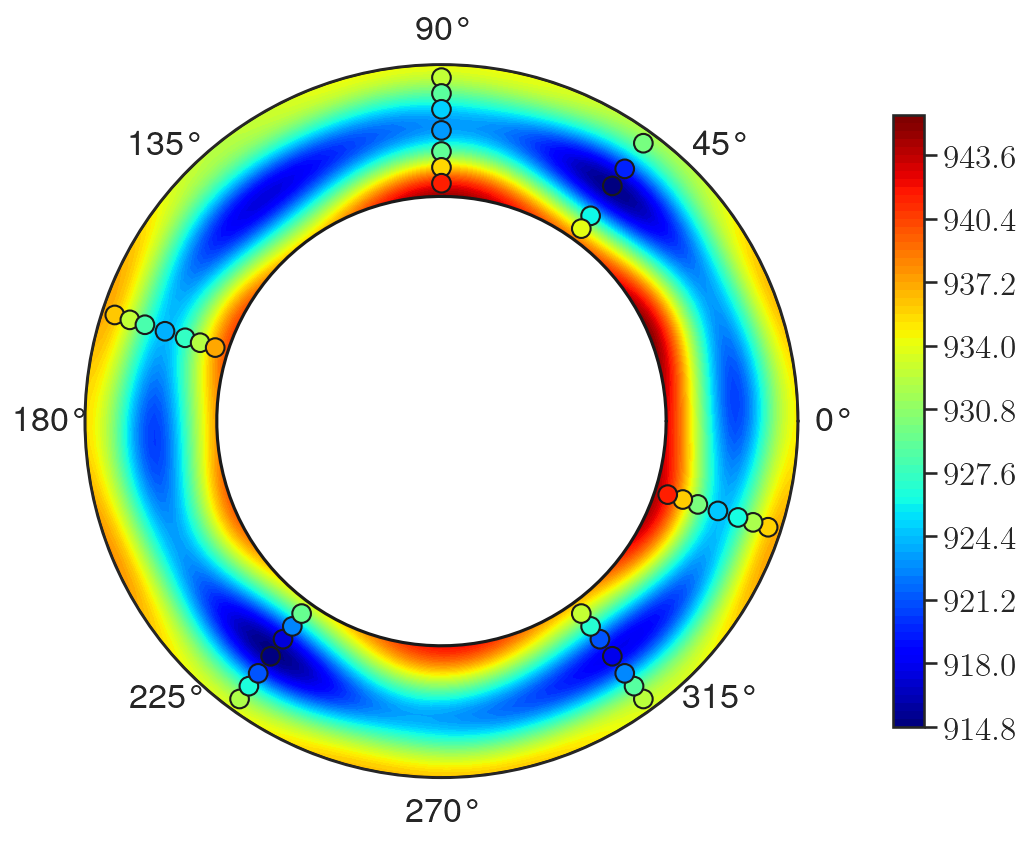}}
\subfigure[]{\includegraphics[]{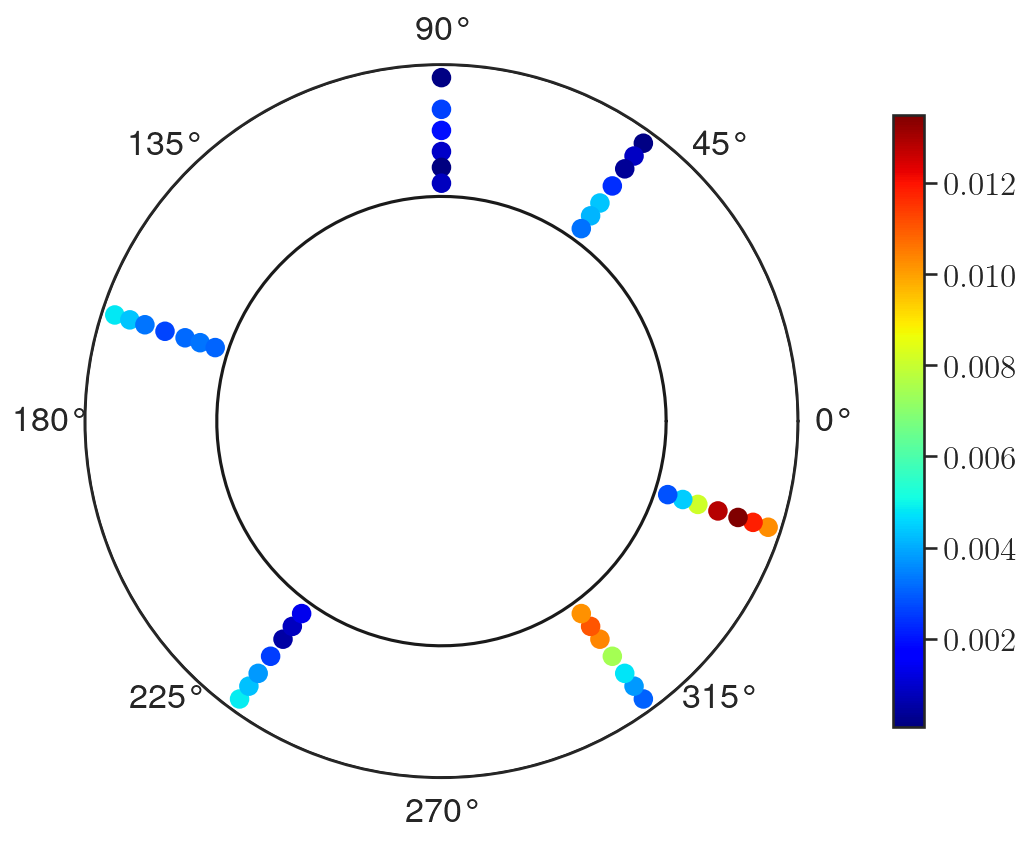}}
\subfigure[]{\includegraphics[]{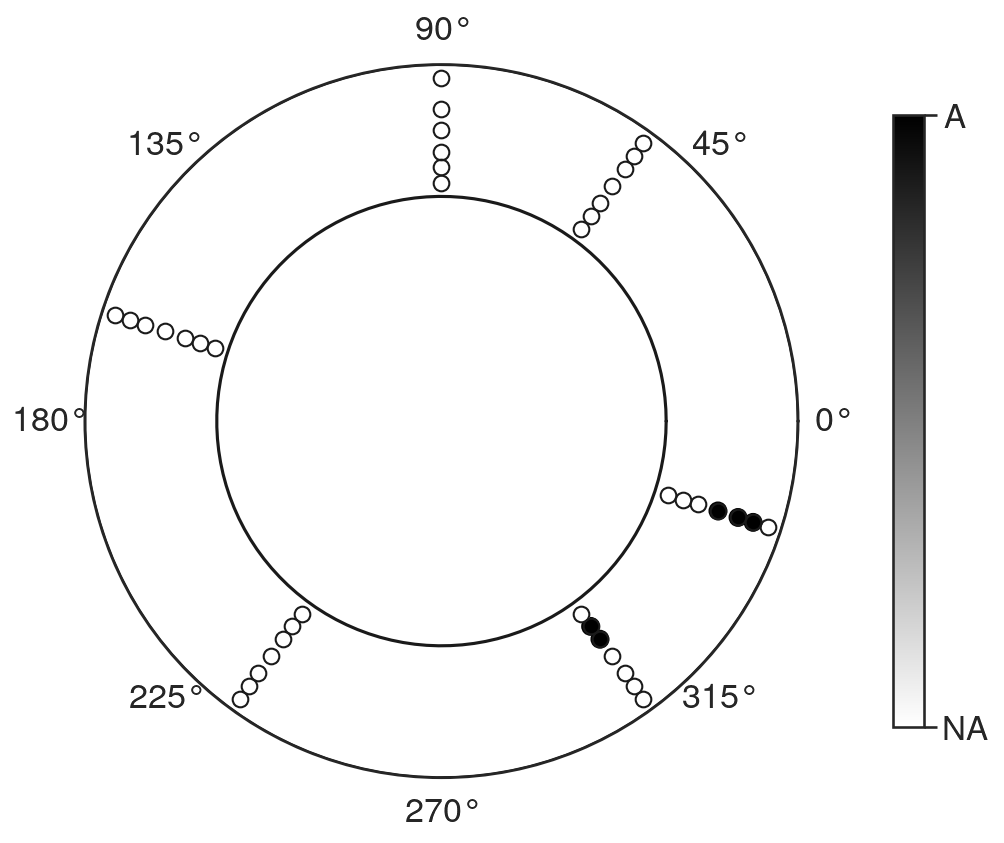}}
\subfigure[]{\includegraphics[]{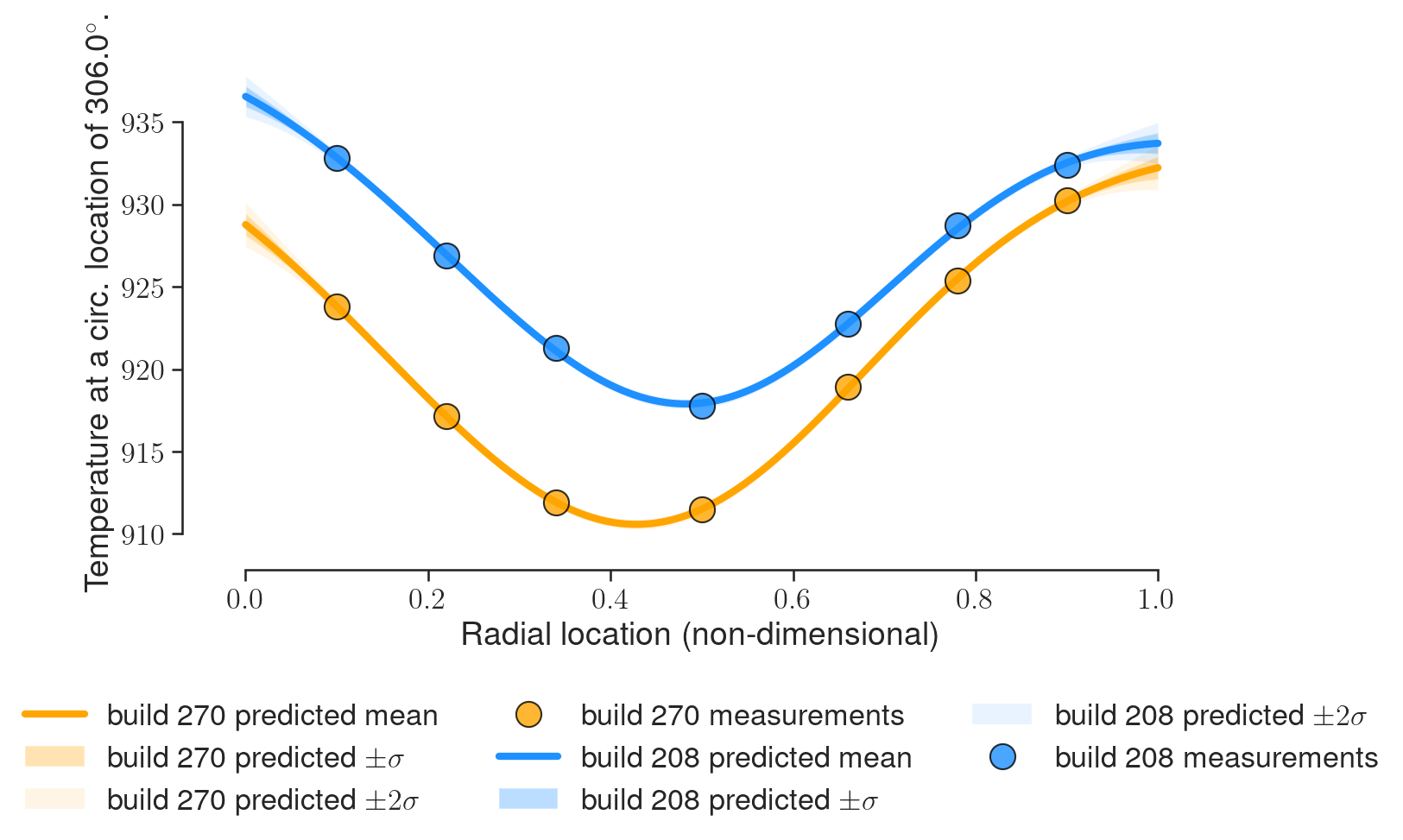}}
\subfigure[]{\includegraphics[]{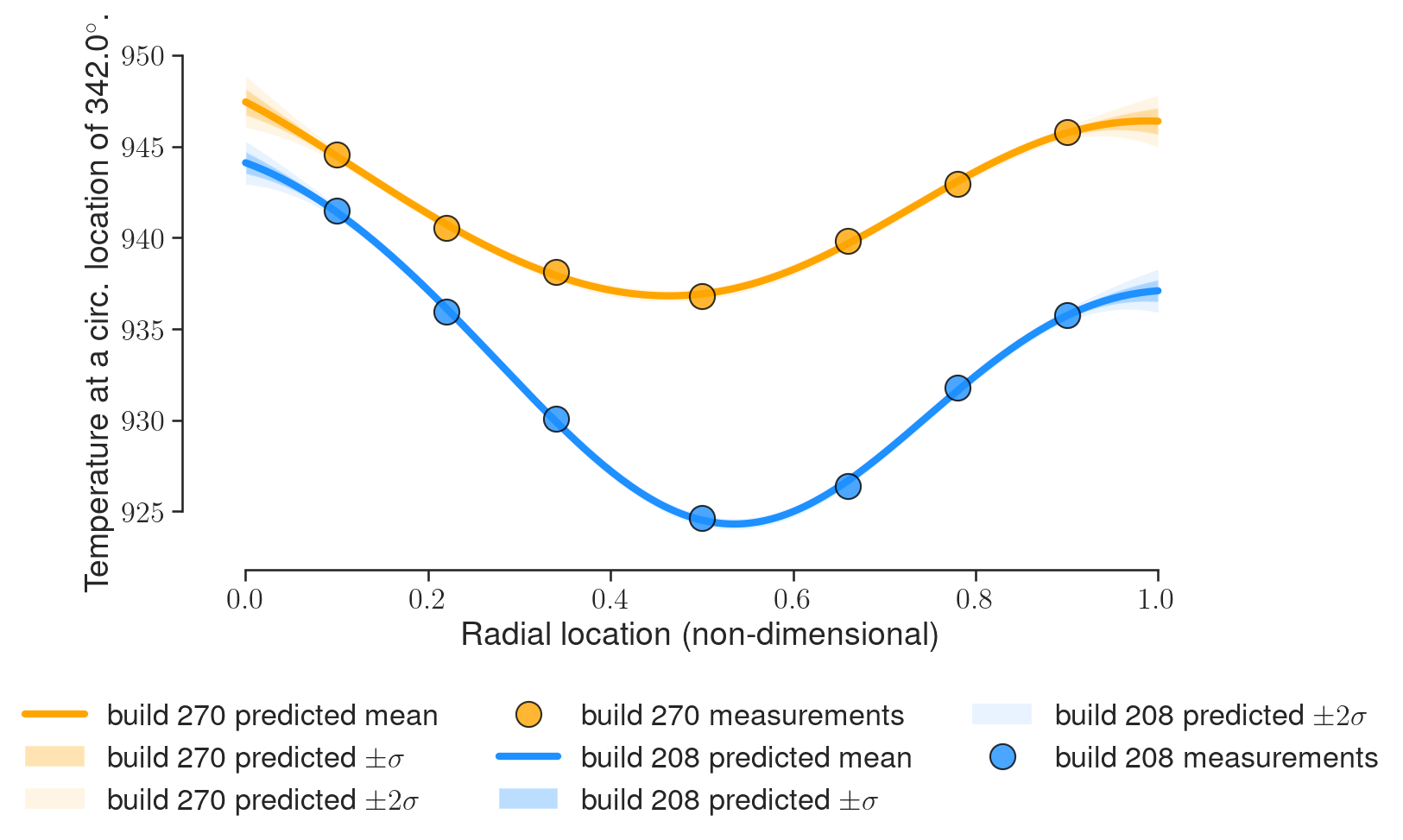}}
\end{subfigmatrix}
\caption{Testing on engine C, station 1: (a) mean of build 270; (b) mean of build 208; (c) distances between the two builds $\vd$ evaluated at the sensor locations of the second build; (d) classified anomalies based on $\tau$ (NA: not anomalous; A: anomalous); (e) radial distribution at $306^{\circ}$, and (f) radial distribution at $342^{\circ}$.}
\label{fig:exposition_station_10}
\end{center}
\end{figure}

The next case studied, also at station 1, is for test asset builds (termed build 270 and 208) from engine project C. Here we have two very distinct builds as is reflected in their spatial mean distributions in Figures~\ref{fig:exposition_station_10}(a) and (b). It is readily apparent that there is something amiss with the rake at $342^{\circ}$ on build 270. Our spatial anomaly detection approach registers this as an anomaly and also picks up an anomaly on the rake at $306^{\circ}$. Radial profiles comparing the two predictive posterior distributions associated with the profiles highlight the extent of dissimilarity. 

\begin{figure}
\begin{center}
\begin{subfigmatrix}{2}
\subfigure[]{\includegraphics[]{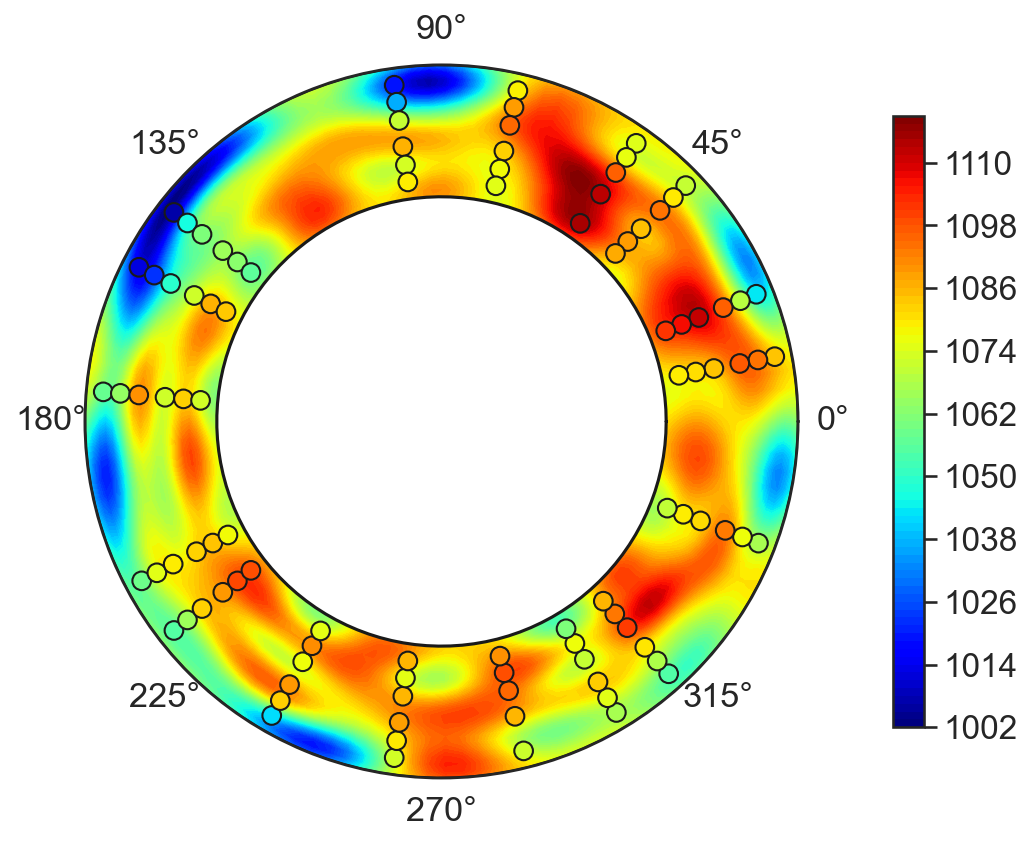}}
\subfigure[]{\includegraphics[]{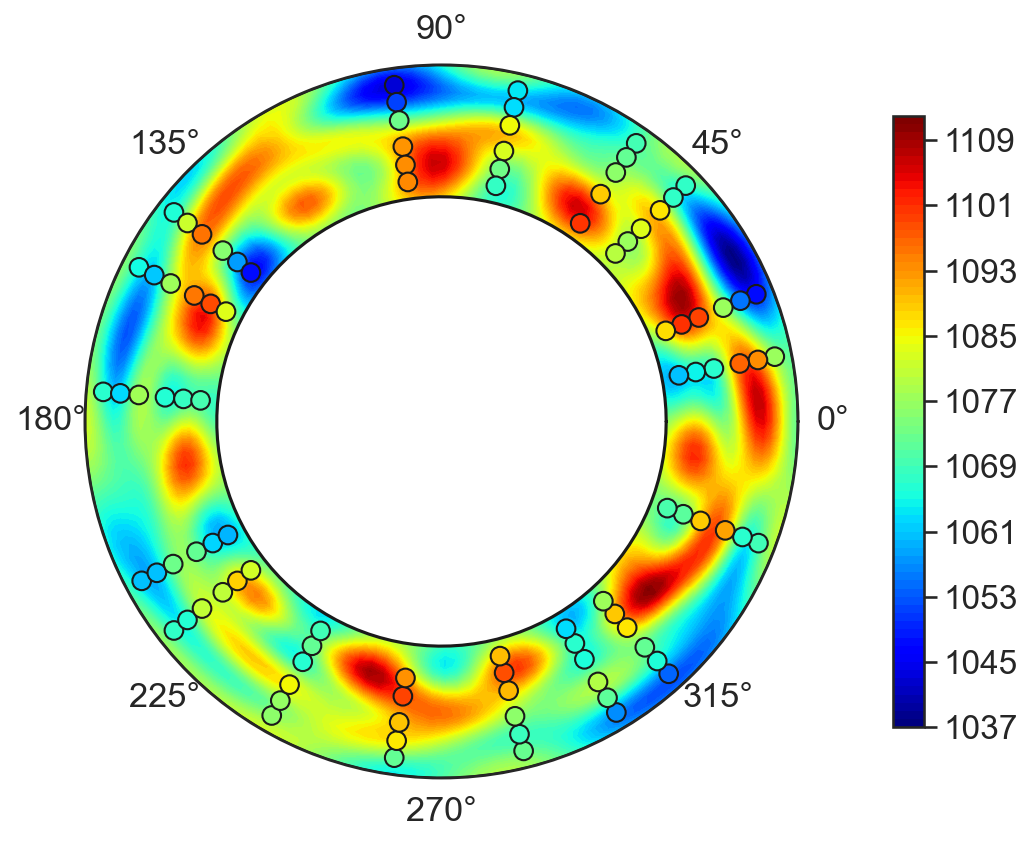}}
\subfigure[]{\includegraphics[]{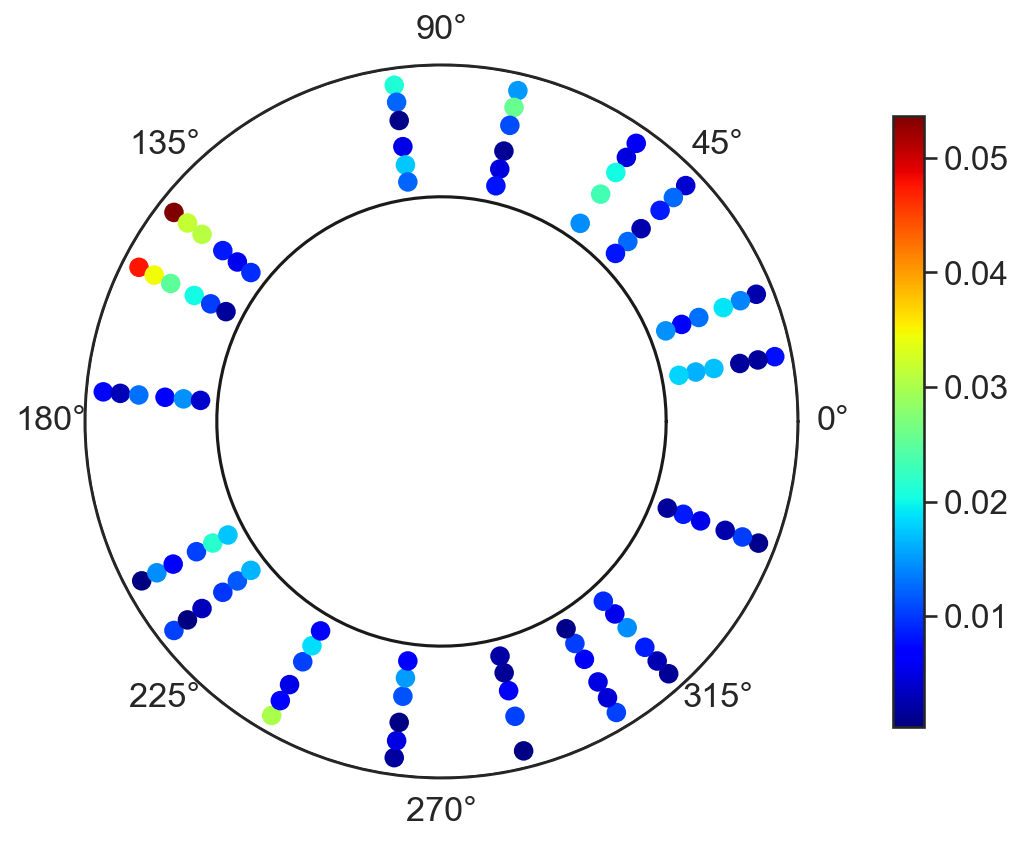}}
\subfigure[]{\includegraphics[]{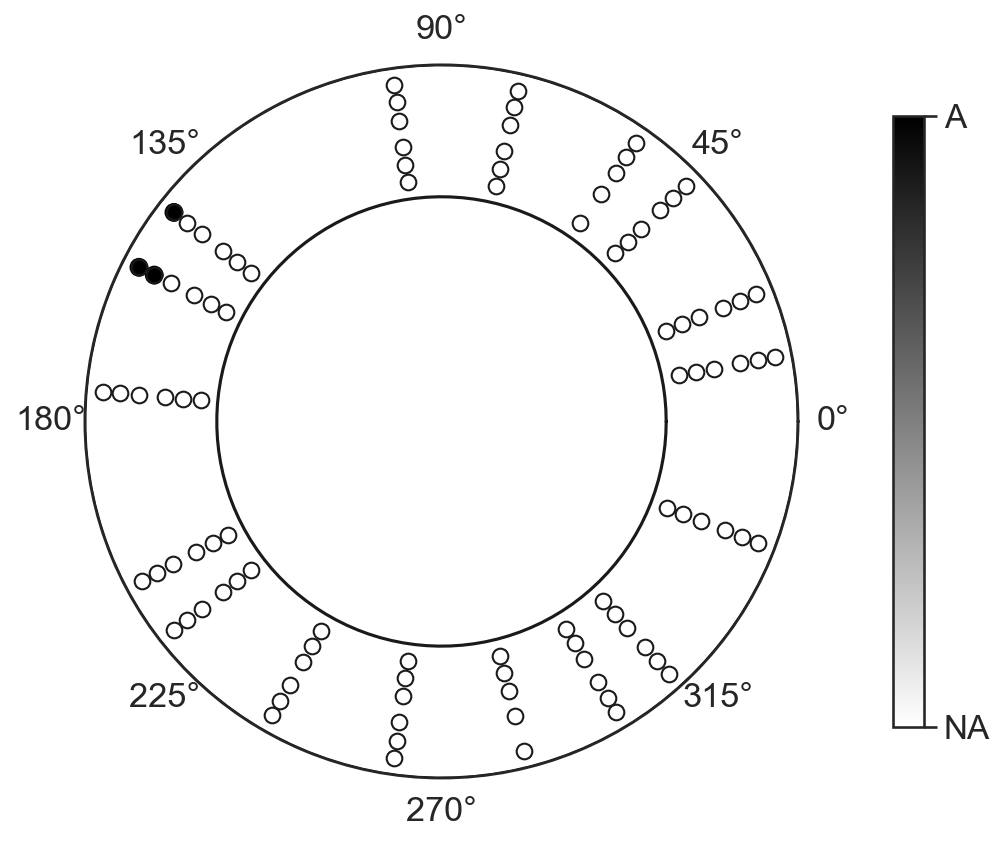}}
\subfigure[]{\includegraphics[]{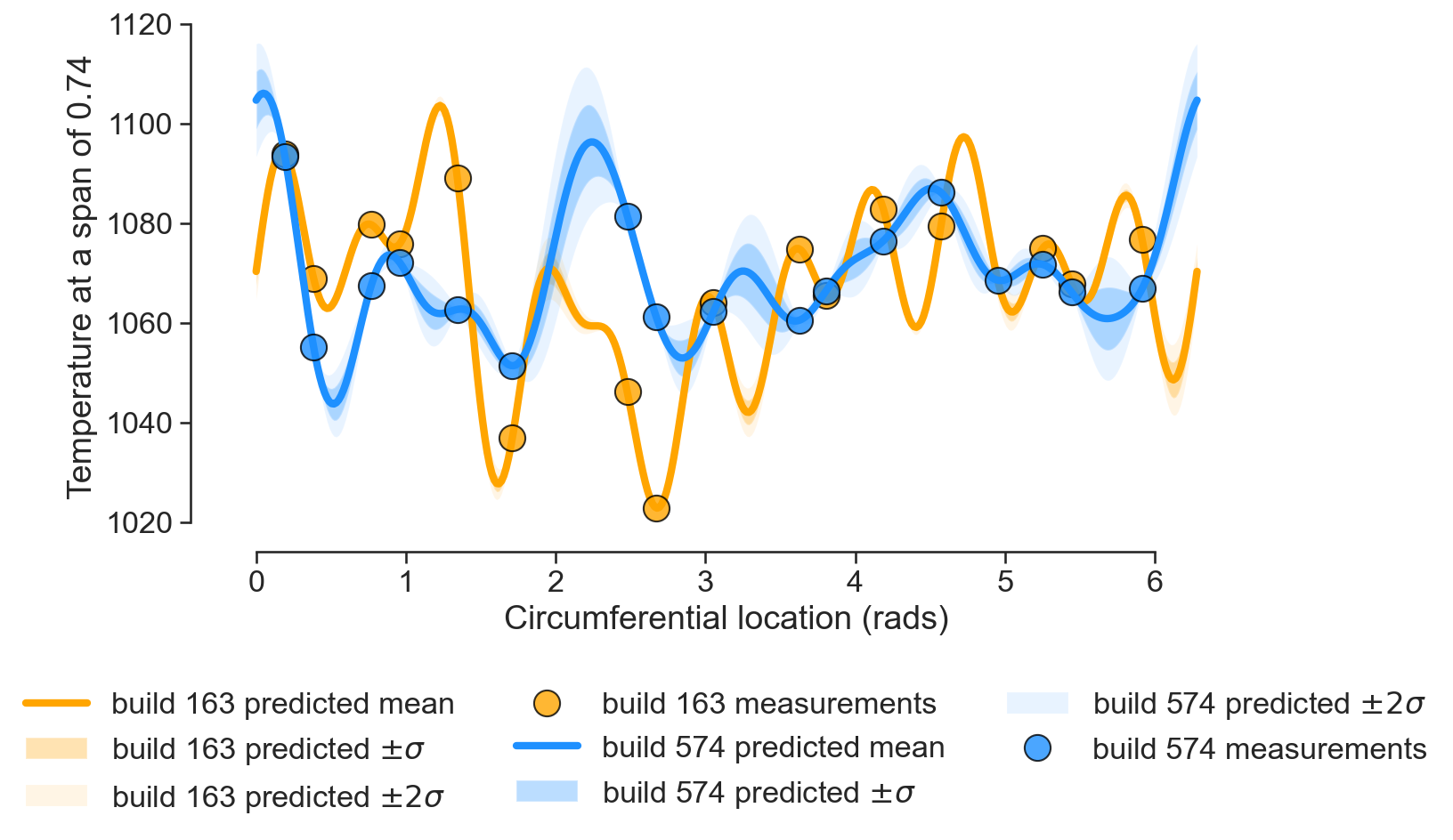}}
\subfigure[]{\includegraphics[]{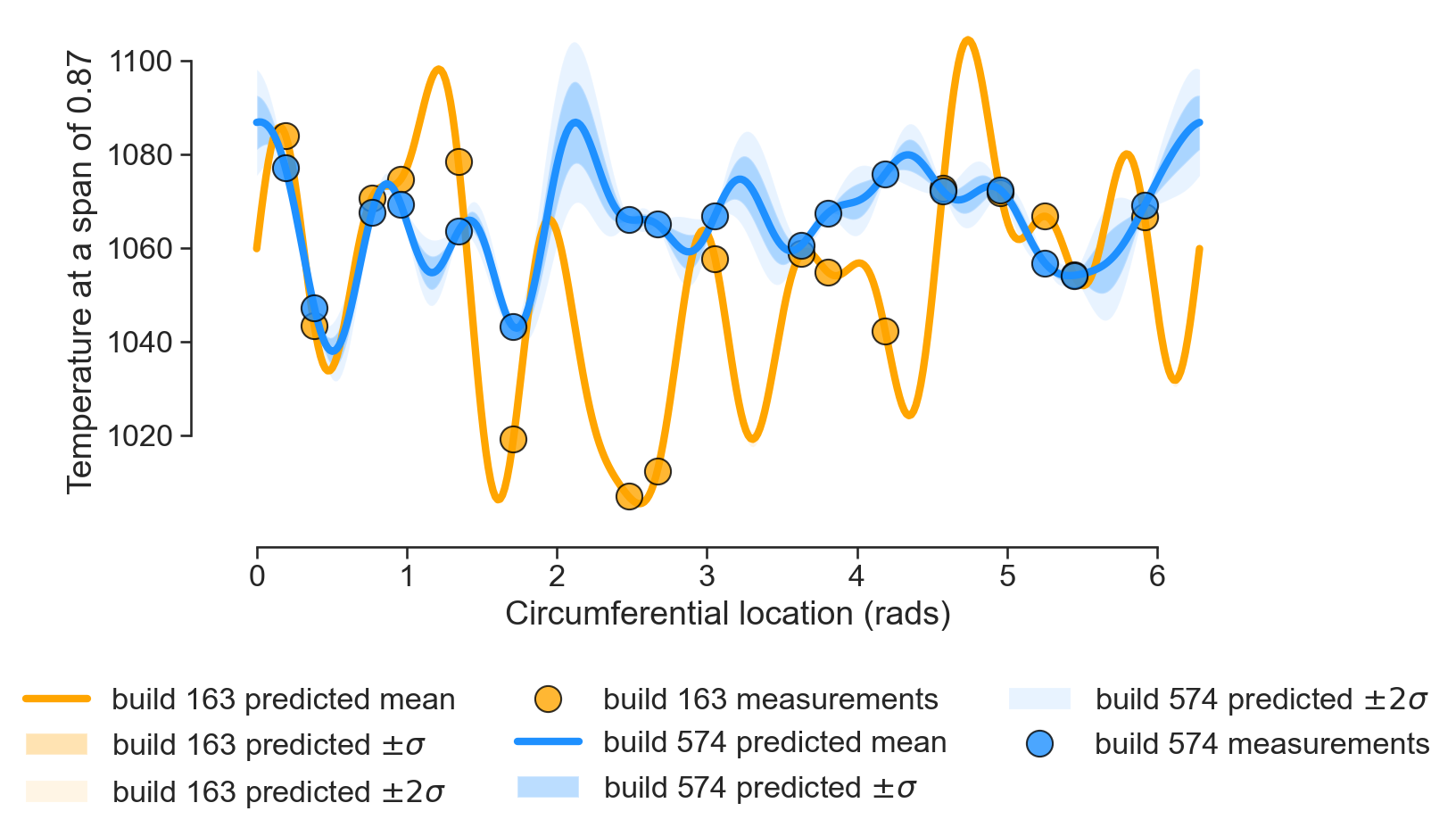}}
\end{subfigmatrix}
\caption{Testing on engine D, station 2: (a) mean of build 163; (b) mean of build 574; (c) distances between the two builds $\vd$ evaluated at the sensor locations of the second build; (d) classified anomalies based on $\tau$ (NA: not anomalous; A: anomalous); (e) circumferential distribution at a span of 0.74, and (f) circumferential distribution at a span of 0.87.}
\label{fig:exposition_station_4}
\end{center}
\end{figure}

\begin{figure}
\begin{center}
\begin{subfigmatrix}{2}
\subfigure[]{\includegraphics[]{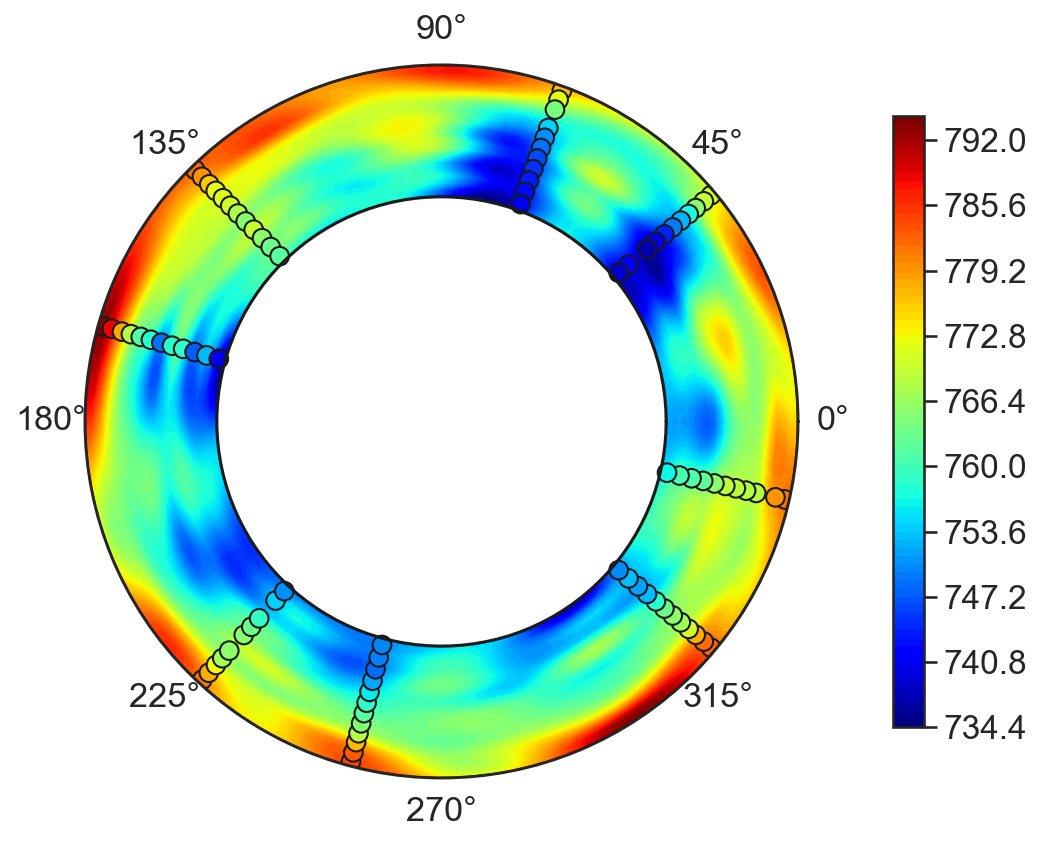}}
\subfigure[]{\includegraphics[]{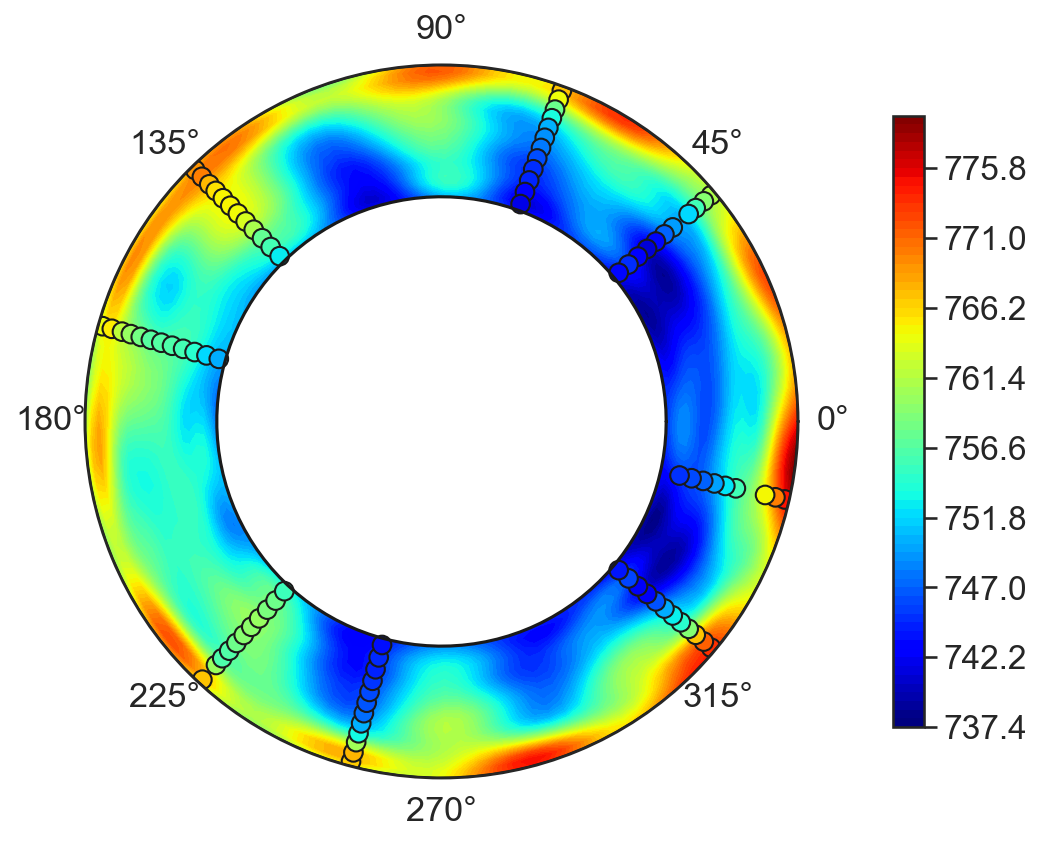}}
\subfigure[]{\includegraphics[]{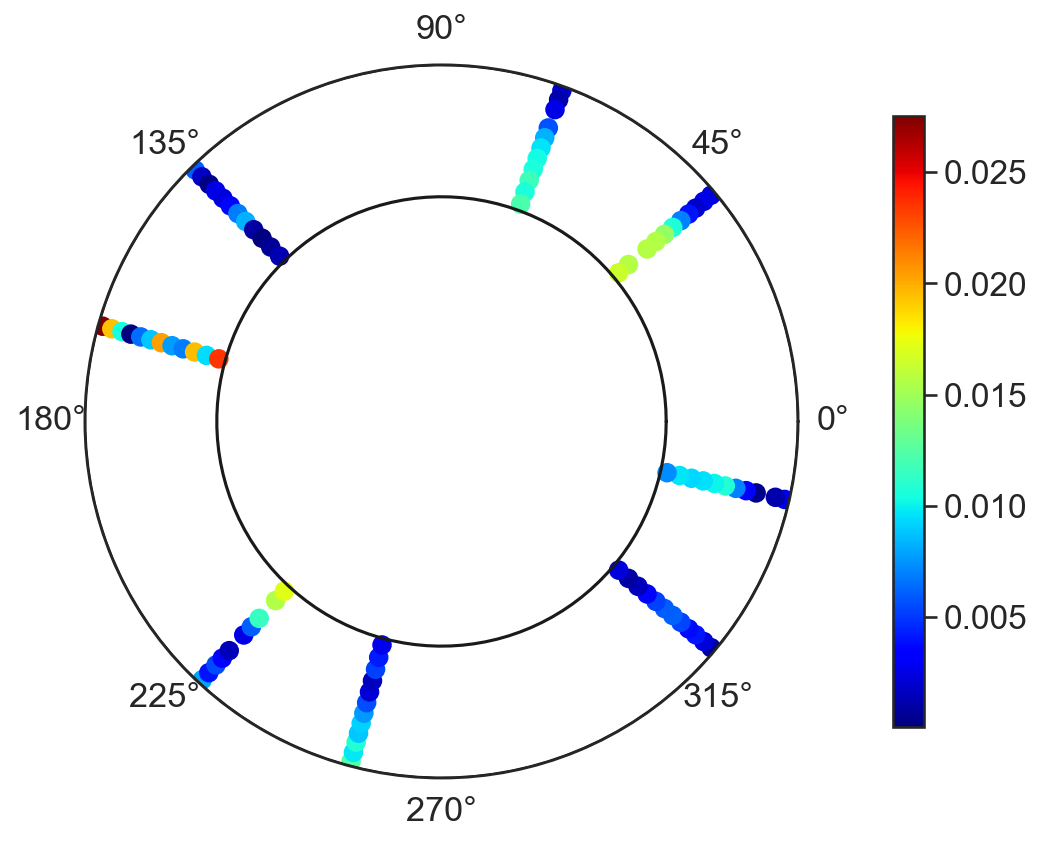}}
\subfigure[]{\includegraphics[]{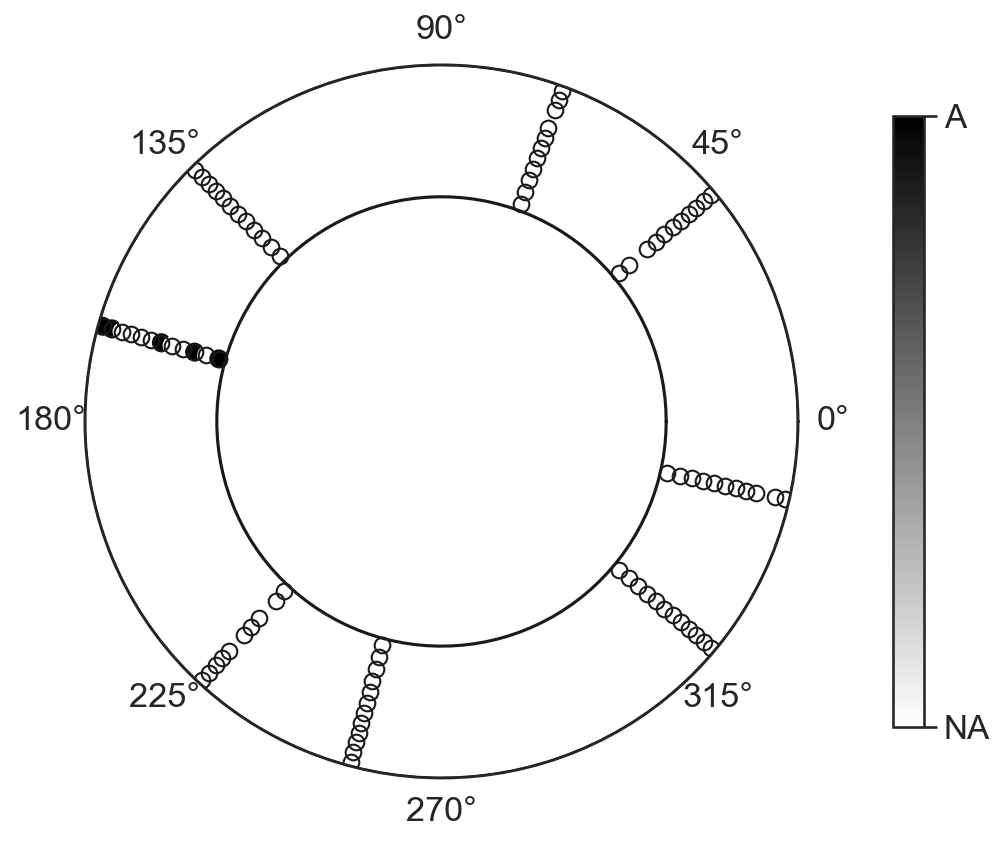}}
\subfigure[]{\includegraphics[]{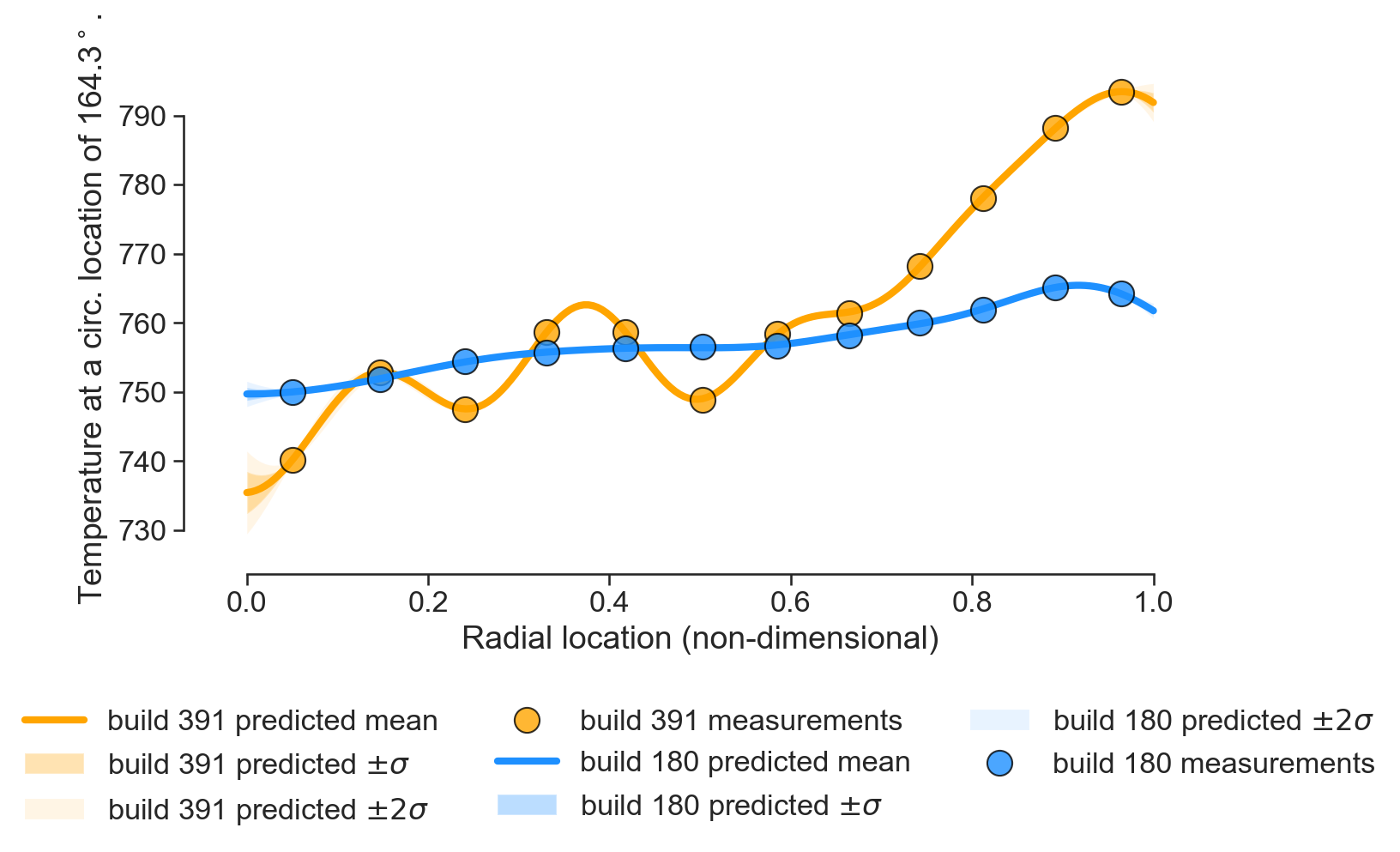}}
\end{subfigmatrix}
\caption{Testing on engine E, station 3: (a) mean of build 391; (b) mean of build 180; (c) distances between the two builds $\vd$ evaluated at the sensor locations of the second build; (d) classified anomalies based on $\tau$ (NA: not anomalous; A: anomalous); (e) radial distribution at $164.3^{\circ}$.}
\label{fig:exposition_station_5}
\end{center}
\end{figure}

Two more such studies are carried out on test asset builds from engine projects D and E in Figures~\ref{fig:exposition_station_4} and \ref{fig:exposition_station_5} and demonstrate the ability of the framework to deal with distinct anomalies. In the case of Figures~\ref{fig:exposition_station_4}, the anomaly was caused by unwanted coolant leakage flow in build 163; in the case of Figures~\ref{fig:exposition_station_5} the culprit was a faulty sensor readings. 

The results above are a snapshot of some of the anomaly detection test cases studied---a selection of a much larger test campaign. While in some cases it is easy to ascertain that a given station has an anomaly via inspection, in most cases it is not. Additionally, there is the time it takes to undertake a manual inspection---having an engineer plot the data for each rake across the distinct measurement stations and comparing the data from an observed engine test to a series of baseline ones---not a matter of minutes. Our framework is fully automated, and as a result drastically reduces the time it takes to identify anomalies. Additionally, it offers a more comprehensive treatment, going beyond the capabilities of a human engineer.  

\subsection{Using the barycenter for anomaly detection}
\begin{figure}
    \centering
    \includegraphics[scale=0.5]{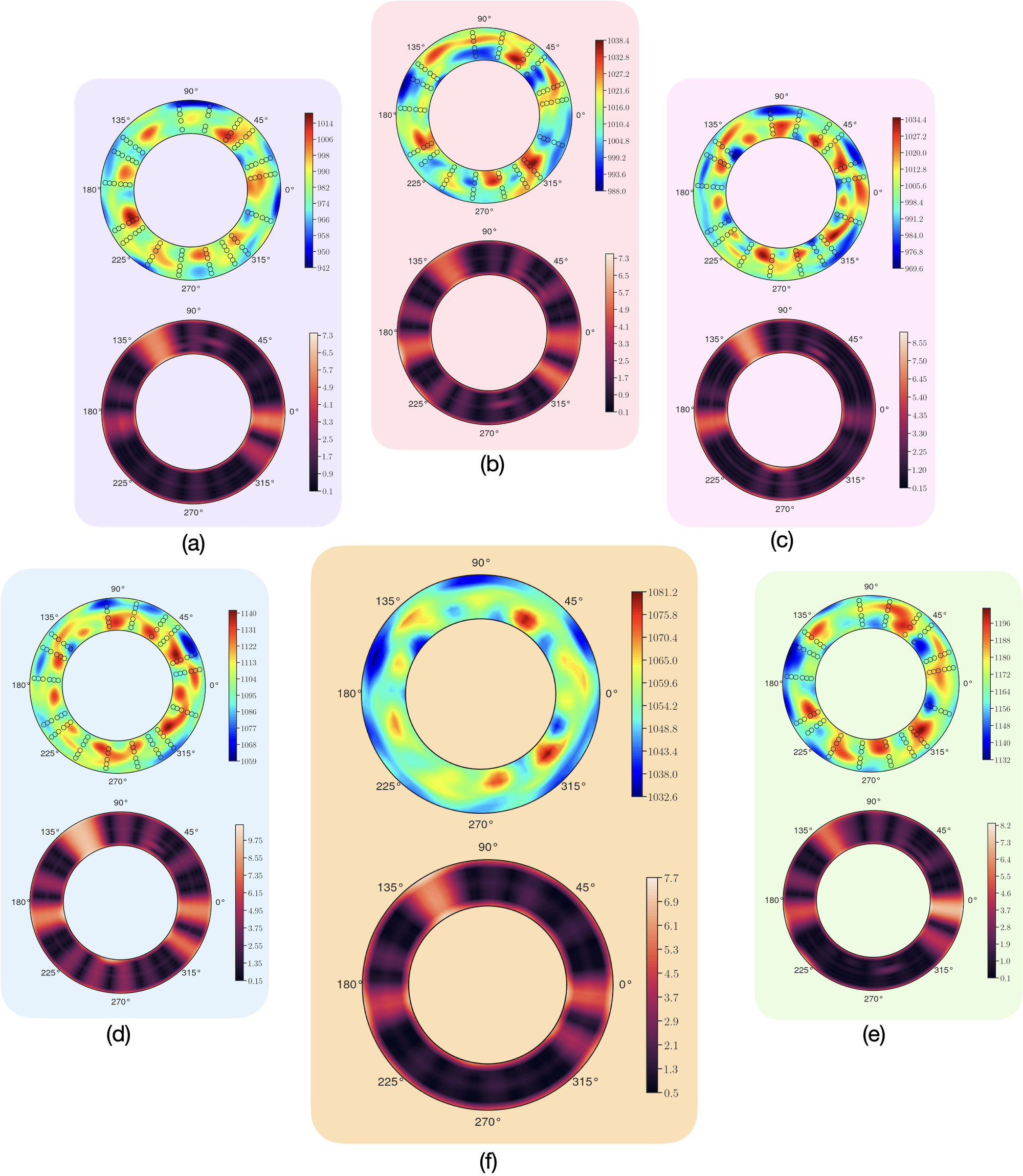}
    \caption{Mean and standard deviations in the predictive posterior distribution for five different builds at station 2 in (a, b, c, d, e). The barycenter is shown in (f). }
    \label{fig:barycenter}
\end{figure}

For completeness, we offer a demonstration of computing the barycenter for multiple distinct distributions at the same plane. Figure~\ref{fig:barycenter}(a-e) shows the mean (top) and standard deviation (bottom) associated with the posterior predictive distributions for five distinct builds at station 2 for engine A. We compute the barycenter \eqref{equ:barycenter_mean_cov} using the fixed point iteration in \eqref{equ:bary_cov_iterative} setting $\vartheta_{j} = 1/5$ for all $j$. Note that in some cases, a weighted barycenter may be more appropriate, i.e., when assigning certain measurement sets more weight than others. Owing to the computational cost of storing and inverting the covariance matrices, we evaluate the barycenter on a coarser grid compared to the distributions above. The final \emph{assimilated} result is shown in Figure~\ref{fig:barycenter}(f). 

\begin{figure}
    \centering
    \includegraphics[scale=0.35]{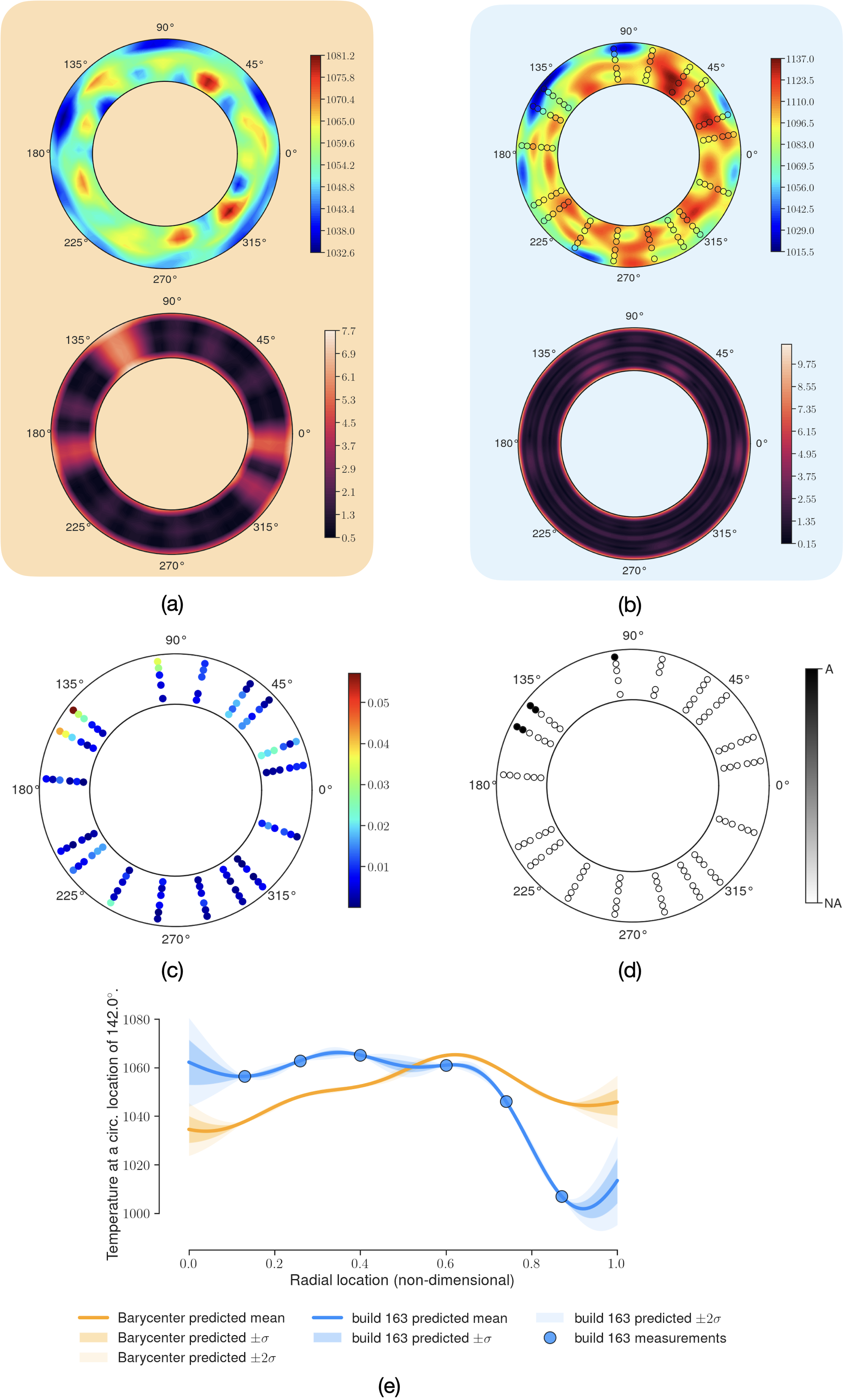}
    \caption{Anomaly detection with the barycenter with mean and standard deviation in (a) and (b) respectively. Build 163's mean and standard deviation are shown in (c) and (d) respectively. Distances between the two cases $\vd$ in (c); classified anomalies based on $\tau$ (NA: not anomalous; A: anomalous) in (d), and radial distribution at $142.0^{\circ}$ in (e).}
    \label{fig:barycenter_2}
\end{figure}

To demonstrate the utility of the barycenter for spatial anomaly detection, we revisit build 163, shown previously in Figure~\ref{fig:exposition_station_4}. Rather than contrast it with data from another build, here we evaluate our anomaly detection approach using the barycenter, with the same threshold $\tau$ determine before. We report the results in Figure~\ref{fig:barycenter_2} and demonstrate that even using the barycenter, the anomalies in build 163 are correctly captured.

\begin{figure}
     \centering
     \includegraphics[scale=0.45]{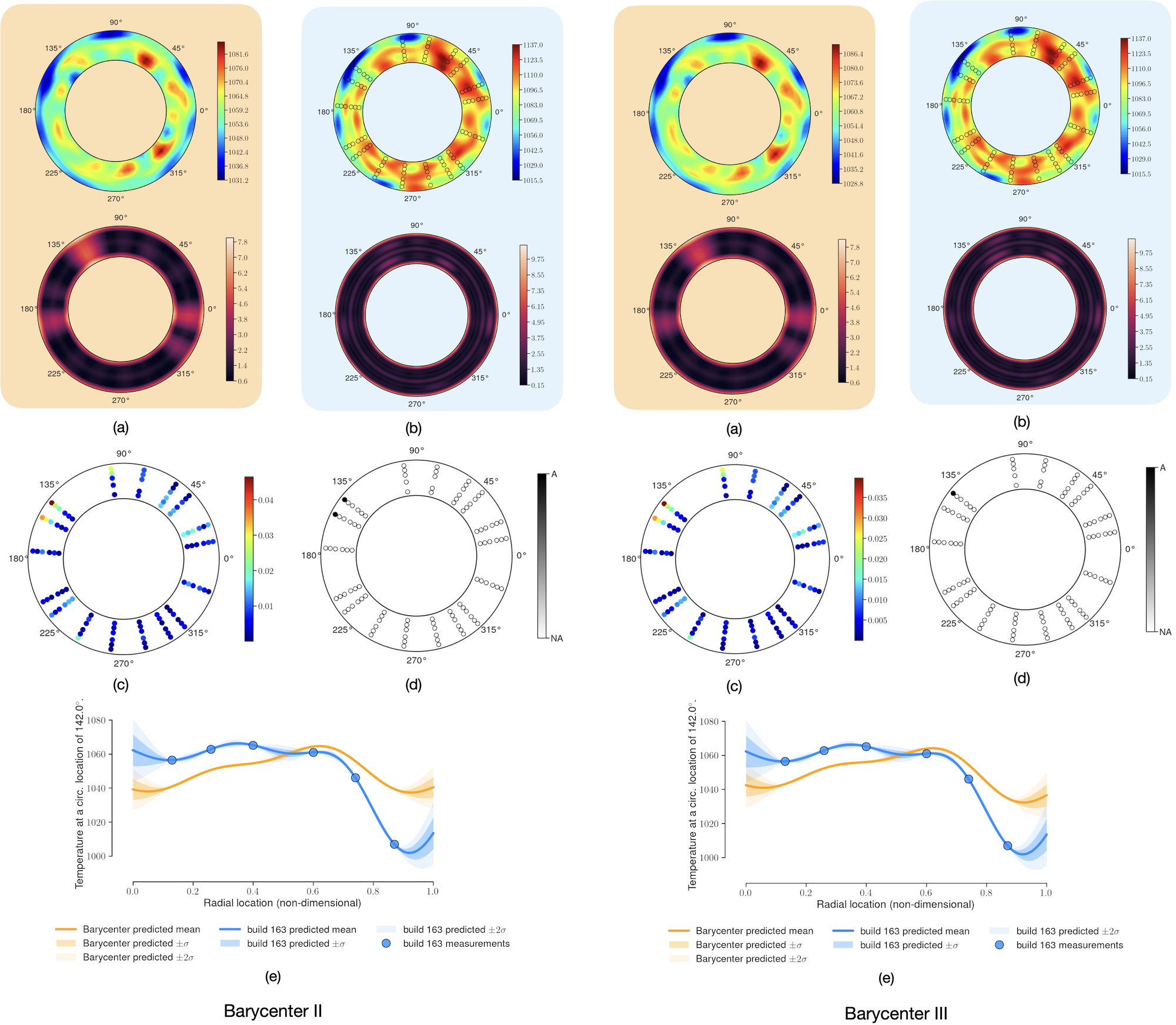}
     \caption{Anomaly detection with barycenters II and III, with mean and standard deviation in (a) and (b) respectively. Build 163's mean and standard deviation are shown in (c) and (d) respectively. Distances between the two cases $\vd$ in (c); classified anomalies based on $\tau$ (NA: not anomalous; A: anomalous) in (d), and radial distribution at $142.0^{\circ}$ in (e).}
     \label{fig:last}
\end{figure}

A valid line of inquiry here is whether a computed barycenter is robust to the inclusion of one or possibly two anomalous data sets. The rationale for this notion is that there may be instances where a given set of measurements may seem non-anomalous until a new build is tested and then compared against. To study this idea, we 
\begin{enumerate}
    \item create a new barycenter for station 2---termed \textbf{barycenter II}---with the five builds in Figure~\ref{fig:barycenter}(a-e) and build 168, and
    \item create another barycenter for station 2---termed \textbf{barycenter III}---with the five aforementioned builds and two instances of build 168. 
\end{enumerate}
The latter is analogous (but not equivalent) to doubling the barycenter weight $\nu$ corresponding to build 168. Then we compute $\vd$ between these new barycenters and build 168; the results are plotted in Figure~\ref{fig:last}. While a clear reduction in the Wasserstein distances are observed, especially for barycenter III, the threshold adequately detects the anomalies, giving us some confidence in this approach.

\section*{Conclusions}
This manuscript presents an anomaly detection framework for identifying spatial anomalies in stagnation temperature measurements in jet engines. It builds upon prior work on interpreting stagnation temperature measurements at an isolated axial plane as a Gaussian random field. We borrow ideas from optimal transport to define a weighted 1D Wasserstein distance between the same locations across two different engine data sets. When this distance exceeds a certain data-driven threshold at an annular location, we classify the corresponding sensor as anomalous. As the definition of an anomaly rests upon what is considered a baseline or gold standard measurement, we exploit the Wasserstein barycenter to aid in assimilating multiple gold standard measurements. 

The results presented in this paper demonstrate the utility of the proposed framework for capturing distinct spatial anomalies. The methodology is invariant to the specific thermodynamic quantity considered, and can also be adapted to other turbomachinery applications.

\section*{Acknowledgements}
The work was part funded by the Fan and Nacelle Future Aerodynamic Research (FANFARE) project under grant number 113286, which receives UK national funding through the Aerospace Technology Institute (ATI) and Innovate UK together with Rolls-Royce plc. The authors are grateful to Rolls-Royce plc for permission to publish this paper. The authors thank Bryn Noel Ubald for his assistance in generating Figure~\ref{fig:ot4}.

\bibliography{references}

\begin{thebibliography}{10}

\bibitem{hipple2020using}
Hipple, S.~M., Bonilla-Alvarado, H., Pezzini, P., Shadle, L., and Bryden,
  K.~M., 2020.
\newblock ``Using machine learning tools to predict compressor stall''.
\newblock {\em Journal of Energy Resources Technology, {\bf 142}}(7),
  p.~070915.

\bibitem{sepe2021physics}
Sepe, M., Graziano, A., Badora, M., Di~Stazio, A., Bellani, L., Compare, M.,
  and Zio, E., 2021.
\newblock ``A physics-informed machine learning framework for predictive
  maintenance applied to turbomachinery assets''.
\newblock {\em Journal of the Global Power and Propulsion Society, {\bf
  2021}}(May), pp.~1--15.

\bibitem{xu2016bayesian}
Xu, S., Jiang, X., Huang, J., Yang, S., and Wang, X., 2016.
\newblock ``Bayesian wavelet pca methodology for turbomachinery damage
  diagnosis under uncertainty''.
\newblock {\em Mechanical systems and signal processing, {\bf 80}}, pp.~1--18.

\bibitem{yan2019accurate}
Yan, W., and Yu, L., 2019.
\newblock ``On accurate and reliable anomaly detection for gas turbine
  combustors: A deep learning approach''.
\newblock {\em arXiv preprint arXiv:1908.09238}.

\bibitem{zhao2016review}
Zhao, N., Wen, X., and Li, S., 2016.
\newblock ``A review on gas turbine anomaly detection for implementing health
  management''.
\newblock In Turbo Expo: Power for Land, Sea, and Air, Vol.~49682, American
  Society of Mechanical Engineers, p.~V001T22A009.

\bibitem{chandola2009anomaly}
Chandola, V., Banerjee, A., and Kumar, V., 2009.
\newblock ``Anomaly detection: A survey''.
\newblock {\em ACM computing surveys (CSUR), {\bf 41}}(3), pp.~1--58.

\bibitem{kullback1951information}
Kullback, S., and Leibler, R.~A., 1951.
\newblock ``On information and sufficiency''.
\newblock {\em The annals of mathematical statistics, {\bf 22}}(1), pp.~79--86.

\bibitem{seshadri2021bayesian}
Seshadri, P., Duncan, A., Thorne, G., Parks, G., Vazquez, R., and Girolami, M.,
  2021.
\newblock ``Bayesian assessments of aeroengine performance with transfer
  learning''.
\newblock {\em arXiv preprint arXiv:2011.14698}.

\bibitem{seshadri2021bayesianmass}
Seshadri, P., Duncan, A., and Thorne, G., 2022.
\newblock ``{Bayesian Mass Averaging in Rigs and Engines}''.
\newblock {\em Journal of Turbomachinery, {\bf 144}}(8), 03.
\newblock 081004.

\bibitem{seshadri2020spatial}
Seshadri, P., Duncan, A., Simpson, D., Thorne, G., and Parks, G., 2020.
\newblock ``Spatial flow-field approximation using few thermodynamic
  measurements—part ii: Uncertainty assessments''.
\newblock {\em Journal of Turbomachinery, {\bf 142}}(2), p.~021007.

\bibitem{seshadri2020spatialb}
Seshadri, P., Simpson, D., Thorne, G., Duncan, A., and Parks, G., 2020.
\newblock ``Spatial flow-field approximation using few thermodynamic
  measurements—part i: Formulation and area averaging''.
\newblock {\em Journal of Turbomachinery, {\bf 142}}(2), p.~021006.

\bibitem{lou2021reconstructing}
Lou, F., and Key, N.~L., 2021.
\newblock ``Reconstructing compressor non-uniform circumferential flow field
  from spatially undersampled data—part 1: Methodology and sensitivity
  analysis''.
\newblock {\em Journal of Turbomachinery, {\bf 143}}(8).

\bibitem{kolouri2017optimal}
Kolouri, S., Park, S.~R., Thorpe, M., Slepcev, D., and Rohde, G.~K., 2017.
\newblock ``Optimal mass transport: Signal processing and machine-learning
  applications''.
\newblock {\em IEEE signal processing magazine, {\bf 34}}(4), pp.~43--59.

\bibitem{schlegl2019f}
Schlegl, T., Seeb{\"o}ck, P., Waldstein, S.~M., Langs, G., and Schmidt-Erfurth,
  U., 2019.
\newblock ``f-anogan: Fast unsupervised anomaly detection with generative
  adversarial networks''.
\newblock {\em Medical image analysis, {\bf 54}}, pp.~30--44.

\bibitem{peyre2019computational}
Peyr{\'e}, G., Cuturi, M., et~al., 2019.
\newblock ``Computational optimal transport: With applications to data
  science''.
\newblock {\em Foundations and Trends{\textregistered} in Machine Learning,
  {\bf 11}}(5-6), pp.~355--607.

\bibitem{agueh2011barycenters}
Agueh, M., and Carlier, G., 2011.
\newblock ``Barycenters in the wasserstein space''.
\newblock {\em SIAM Journal on Mathematical Analysis, {\bf 43}}(2),
  pp.~904--924.

\bibitem{alvarez2016fixed}
{\'A}lvarez-Esteban, P.~C., Del~Barrio, E., Cuesta-Albertos, J., and
  Matr{\'a}n, C., 2016.
\newblock ``A fixed-point approach to barycenters in wasserstein space''.
\newblock {\em Journal of Mathematical Analysis and Applications, {\bf
  441}}(2), pp.~744--762.

\bibitem{dyer2015riemannian}
Dyer, R., Vegter, G., and Wintraecken, M., 2015.
\newblock ``Riemannian simplices and triangulations''.
\newblock {\em Geometriae Dedicata, {\bf 179}}(1), pp.~91--138.

\end{thebibliography}
\bibliographystyle{asmems4}

\end{document}